\documentclass[12pt]{article}

\usepackage{amsmath}
\usepackage{amssymb}
\usepackage{graphics}
\usepackage{graphicx}
\usepackage{multirow}
\usepackage{bm}
\usepackage{lscape}  
\usepackage[authoryear]{natbib}
\usepackage{color}
\usepackage{xcolor}
\usepackage{framed}
\usepackage{geometry}
\usepackage{lineno}   % for line numbers
\usepackage{setspace}
\usepackage{fancyhdr}
\usepackage{textcomp}
\usepackage{float}
\usepackage{tabularx}
\usepackage{booktabs}
\usepackage{caption}
\definecolor{fgcolor}{rgb}{0.345, 0.345, 0.345}
\definecolor{shadecolor}{rgb}{.97, .97, .97}
\definecolor{messagecolor}{rgb}{0, 0, 0}
\definecolor{warningcolor}{rgb}{1, 0, 1}
\definecolor{errorcolor}{rgb}{1, 0, 0}
\definecolor{azul}{RGB}{0, 128, 255} % color for the links

\makeatletter
\def\maxwidth{ %
  \ifdim\Gin@nat@width>\linewidth
    \linewidth
  \else
    \Gin@nat@width
  \fi
}
\makeatother

\makeatletter
 {\par\unskip\endMakeFramed%
 \at@end@of@kframe}
\makeatother
\usepackage[unicode=true,pdfusetitle,
 bookmarks=true,bookmarksnumbered=true,bookmarksopen=true,bookmarksopenlevel=2,
 breaklinks=false,pdfborder={0 0 1},backref=false,colorlinks=true]{hyperref}
 \hypersetup{
  citecolor= darkgray,%red,
  linkcolor=azul,
  urlcolor=azul,
 pdfstartview={XYZ null null 1},
 linktoc=all    % color the numbers in the toc
 }
 
 \def \R{\mathrm{I}\!\mathrm{R}}

\begin{document}

%\linenumbers
% title and authors
\doublespacing
\begin{center}
{\LARGE{\bf 
Prediction Measures in Nonlinear Beta Regression Models}}\\ %A comparison of various estimators for the coefficient of variation}}\\
\vspace{.7cm}
\doublespacing
\large
Patr\'{\i}cia Leone  Espinheira$^{1}$ \hspace{.2cm} 
Luana C. Meireles da Silva$^{1}$ \hspace{.2cm} 
Alisson de Oliveira Silva$^{1}$ \hspace{.2cm}
Raydonal Ospina$ ^{1, \#}$
\end{center}

%\vspace{.1cm}
%\begin{center}
%%\today%This version: \today \\
%%First version: February 23, 2016
%\end{center}

%\doublespacing
\singlespacing

% afiliations
%\vspace{.1cm}
%{\it 
%\noindent $^1$  Universidad del Norte, Barranquilla, Colombia. \\
%\noindent $^2$  The Arcos-Burgos Group, Department of Genome Sciences, John Curtin School of Medical Research, The Australian National University, Canberra, ACT, Australia. \\
\noindent $^1$  {Departamento de Estat\'{\i}stica, Universidade Federal de Pernambuco, Cidade Universit\'aria, Recife/PE, 50740--540, Brazil. \\
%\noindent $^4$  G\"osta Ekman Laboratory,
%                Department of Psychology,
%                Stockholm University, Stockholm, Sweden.\\
%\noindent $^5$  School of Psychology, The University of Adelaide, Adelaide, Australia.\\                 
\noindent $^\#$  E-mail:} 
 \href{mailto:raydonal@de.ufpe.br}{\texttt{raydonal@de.ufpe.br}}
% } 

\singlespacing

% ------------------------------
%          Resumen
% ------------------------------
\begin{abstract}
\normalsize
\noindent Nonlinear models are frequently applied to determine the optimal supply natural gas to a given residential unit based on economical and technical factors, or used to fit biochemical and pharmaceutical assay nonlinear data. 
In this article we propose PRESS statistics  and prediction coefficients  for a class of nonlinear beta regression models, namely $P^2$ statistics. We aim at using both prediction coefficients and goodness-of-fit  measures as a scheme of  model select criteria. In this sense, we introduce for beta regression models under nonlinearity the use of the model selection criteria based on robust pseudo-$R^2$ statistics.
% $R^2_{LR}$ and $R^2_{FC}.$ 
%(\cite{BAYER2017}).  
Monte Carlo simulation results on the finite sample behavior of both prediction-based
model selection criteria $P^2$ and the pseudo-$R^2$ statistics are provided. 
Three applications for real data are presented.
The linear application  relates  to the distribution
of natural gas for home usage in S\~ao Paulo, Brazil. Faced with the economic risk of too overestimate or to underestimate the distribution of gas has been necessary to construct prediction limits  
and to select the best predicted and fitted model to construct best prediction limits it is the aim of the first application. Additionally, the two nonlinear applications presented also highlight the importance of considering  both goodness-of-predictive and goodness-of-fit of the competitive models.
\end{abstract}

%% text at the top left
%\lhead{\textit{A robust standard measure of data's dispersion}}
%% text at the top right
%\rhead{\textit{V\'elez, Ospina \& Marmolejo-Ramos, 2016}}
%% text at the bottom
\cfoot{\thepage}   % page numbers

\singlespacing

% ---------- Document ------------------------------

\section{Introduction}

 \cite{Ferrari2004} introduced a regression model in which the response is beta-distributed, its mean being related to a linear predictor through a link function. The linear predictor includes independent variables and regression parameters. Their model also includes a precision parameter whose reciprocal can be viewed as a dispersion measure. In the standard formulation of the beta regression model it is assumed that the precision is constant across observations. However, in many practical situations this assumption does not hold. \cite{Smithson2006}  consider a beta regression specification in which dispersion is not constant, but is a function of covariates and unknown parameters and  \citep{SIMAS2010} introduces the class of nonlinear beta regression models. Parameter estimation is carried out by  maximum likelihood (ML) and standard asymptotic hypothesis testing can be easily performed. Practitioners can use the {\tt betareg} package, which is available for the {\tt R} statistical
software (\url{http://www.r-project.org}), for fitting beta regressions.  \cite{Cribari2010} provide an overview of varying dispersion beta regression modeling using the {\tt betareg} package. Diagnostic tools and improve ML estimation were accomplished in \cite{Espinheira2008,Espinheira+Ferrari+Cribari_2008b,Ospina2006960,Chien2011} and others. Inference for beta regression also have been  developed in an  Bayesian context (\cite{Vale+Ferrari+Zuniga_2013,Brascom+Johnson+Thurmond_2007} and \cite{Cepeda+Gamerman_2005}.)

Recently \cite{Espinheira2014} build and evaluated bootstrap-based prediction intervals for the class of beta regression models with varying dispersion. However, a prior approach it is necessary, namely: the selection of 
the model with the best predictive  ability, regardless of the goodness-of-fit. Indeed, the model selection is a crucial step in data analysis, since all inferential performance is based on the selected model.  \cite{BAYER2017} evaluated the performance of different model selection criteria in samples of finite size in a beta regression model, such as Akaike Information Criterion (AIC) \citep{Aka:1973}, Schwarz Bayesian Criterion (SBC) \citep{Schwarz78} and  various approaches based on pseudo-$R^2$ such as the coefficient of determination adjusted $R^2_{FC}$ proposed by \cite{Ferrari2004} and the version based on log-likelihood functions, namely by $R^2_{LR}.$ Indeed, the authors proposed two new model selection criteria and a fast two step model selection scheme considering both mean and dispersion submodels, for  beta regression models with varying dispersion.

However, these methods  do not offer any insight about the quality of the predictive values in agreement with the findings of \cite{Spiess2010} for nonlinear models. 
In this context, \cite{All:1974}, proposed the PRESS (Predictive Residual Sum of Squares) criterion, that can be used as a measure of the predictive power of a model.  The PRESS statistic  is independent from the goodness-of-fit  of the model, since, that its calculation is made by leaving out  the observations  that the model  is trying to predict \citep{Palmer}. The PRESS statistics can be viewed as a sum of squares of external residuals \citep{Bartoli2009}. Thus, similarly of the approach of $R^2,$  \cite{Mediavilla2008}  proposed a coefficient of prediction based on PRESS namely $P^2$.  The  $P^2$ statistic can be used to select models from a predictive perspective adding important information about the predictive ability of the model in various scenarios.  

Our chief goal in this paper is to propose
versions of the PRESS statistics and the coefficients of prediction $P^2$ associated, for the linear and nonlinear beta regression models. As a second contribution, in especial to beta regression under nonlinearity, we evaluate the behavior of $R^2_{LR}$\citep{BAYER2017} and $R^2_{FC}$\citep{Ferrari2004} measures both when the model is correctly specified and when under model misspecification. The results of the simulations showed as the prediction coefficients can be useful in detecting misspecifications, or indicate difficulties on to estimate beta regression models when the data are close to the boundaries of the standard unit interval. Finally, the real data applications are the last and important contribution. Here we provide guidance for researchers in choosing and interpreting the measures proposed.  In fact, based on these applications we can shown how it is important to consider both coefficients of prediction and  coefficients of determination  to  build models more useful to describe the data.
%
%Based on these applications we can shown, in fact, how it is important to consider both coefficients of prediction and  coefficients of determination  to  build models more useful to describe the data.

\section{The $P^2$ statistic measure}\label{sec3}
Consider the linear model,
$Y = X\beta + \varepsilon$
\noindent where $Y$ is a vector $n \times 1$ of  responses, $X$ is a known matrix of covariates of dimension $n \times p$ of full rank, $\beta$ is the parameter vector of dimension $p \times 1$  and $\varepsilon$ is  a vector  $n \times 1$ of errors.  We have the least squares estimators: $\widehat{\beta} = (X^{\top}X)^{-1}X^{\top}y$, the residual $e_t = y_t - x_t^{\top}\widehat{\beta}$ and the predicted value $\widehat{y}_t = x_t^{\top}\widehat{\beta},$ where $x^\top_t=(x_{t1}, \ldots, x_{tp}),$ and $t=1,\ldots,n$.   Let $\widehat{\beta}_{(t)}$ be the least squares estimate of $\beta$ without the $t$th observation and $\widehat{y}_{(t)} = x_t^{\top}\widehat{\beta}_{(t)}$ be the predicted value of the case deleted, such that $e_{(t)} = y_t - \widehat{y}_{(t)}$ is the prediction error or external residual. Thus, for multiple regression, the classis statistic
\begin{equation}
\label{press:1}
PRESS = \sum_{t = 1}^{n}e_{(t)}^2 = \sum_{t = 1}^{n}(y_t - \widehat{y}_{(t)})^2,
\end{equation}
 which can be rewritten as $PRESS = \sum_{t = 1}^{n}(y_t - \widehat{y}_t)^2/(1 - h_{tt})^2$, where $h_{tt}$ is the t$th$ diagonal element of the projection matrix $X(X^{\top}X)^{-1}X^{\top}$.

Now, let $y_1, \ldots, y_n$ be independent random variables such that each $y_t$, for $t=1,\ldots,n$, is beta distributed, beta-distributed, denoted by $y_t \sim{\cal B}(\mu_t, \phi_t)$,
i.e., each $y_t$ has density function given by
\begin{equation}\label{eq1}
	f(y_t; \mu_t, \phi_t) = \frac{\Gamma(\phi_t)}{\Gamma(\mu_t\phi_t) 
		\Gamma((1-\mu_t)\phi_t)} y_t^{\mu_t\phi_t-1}(1-y_t)^{(1-\mu_t)\phi_t-1},\quad
	0 < y_t < 1,
\end{equation}
where $0 < \mu_t < 1$ and $\phi_t > 0$.
Here, ${\rm E}(y_t) = \mu_t$ and ${\rm Var}(y_t) = {V(\mu_t) / (1+\phi_t)}$, where $V(\mu_t) = \mu_t(1-\mu_t)$.
\cite{SIMAS2010} proposed the class of nonlinear beta regression models in which the mean of $y_t$ and the precision parameter can be written as  
\begin{equation}
\label{component}
	g(\mu_t)=\eta_{1t}=f_1(x^\top_t;\beta) \quad \mbox{and} \quad h(\phi_t)=\eta_{2t}=f_2(z^\top_t,\gamma),
\end{equation}
where $\beta = (\beta_1, \ldots, \beta_k)^{\!\top}$ and $\gamma = (\gamma_1, \ldots, \gamma_q)^{\!\top}$
are, respectively, $k\times 1$ and $q\times 1$ vectors of unknown parameters
($\beta \in \R^k$; $\gamma \in \R^q$),
$\eta_{1t}$ and $\eta_{2t}$ are the nonlinear predictors,  $x^\top_t=(x_{t1}, \ldots, x_{tk_1})$ and $z^\top_t=(z_{t1}, \ldots, z_{tq_1})$ are vectors of covariates (i.e., vectors of independent variables),  $t=1, \ldots, n$, $k_1 \leq k$, $q_1 \leq q$ and $k+q< n$. 
Both $g(\cdot)$ and $h(\cdot)$ are strictly monotonic
and twice differentiable link functions. Furthermore, $f_i(\cdot)$, $i=1, 2$, are differentiable and continous functions, such that  
the matrices $J_1=\partial \eta_1 / \partial \beta$ and $J_2=\partial \eta_2 / \partial \gamma$ have full rank (their ranks are equal to $k$ and $q$, respectively). The parameters that index the model can be estimated by maximum likelihood (ML). In the Appendix, we present the log-likelihood function, the score vector and Fisher's information matrix for the nonlinear beta regression model.

In the nonlinear beta  regression model, the ML estimator $\widehat \beta$  can be viewed as the least squares estimator of  $\beta$ (see Appendix) obtained
by regressing 
\begin{equation}\label{eq2}
\check{y} =\widehat\Phi^{1/2} {\,\widehat{\! W}}^{1/2}u_1\,\,\,  \text {on} \,\,\check{J_1} = \widehat\Phi^{1/2}{\,\widehat{\! W}}^{1/2}J_1,\end{equation}
with $\Phi = {\rm diag}(\phi_1,\ldots,\phi_n)$, $J_1=\partial \eta_{1}/\partial \beta.$ Here,  matrix $W$ and $u_1$ are given in \eqref{A3}--\eqref{A5} in the Appendix.  
%
%defined in \eqref{A5} and $t$th element of the diagonal matrix $W$ being given in \eqref{A3}.
Thus, the prediction error is
$\check{y}_t - \widehat{\check{y}}_{(t)} = \widehat\phi_t^{1/2}\widehat{w}^{1/2}_t u_{1,t} - \widehat\phi_t^{1/2}\widehat{w}^{1/2}_t J_{1t}^{\top}\widehat{\beta}_{(t)}$, in which $J_{1t}^{\top}$ is the $t$th row of the $J_1$ matrix. Using the ideas proposed by \citep{pregibon1981} we have that 
 $\widehat{\beta}_{(t)} = \widehat{\beta} - \{{(J_1^{\top}\widehat \Phi{\,\widehat{\! W}}J_1)^{-1}J_{1t} \widehat\phi_t^{1/2}{\,\widehat{\! w}}_t^{1/2} r^{\beta}_t}\}/({1 - h_{tt}^*}),$  
 where $r^{\beta}_t$ is the weighted $1$ residual \citep{Espinheira2008} defined as 
 \begin{equation}\label{eq3}
 	\begin{split}
 		r^\beta_{t}=\frac{y^*_t-\widehat{\mu}^*_t}{\sqrt{\widehat{v}_t}}, 
	\end{split} 
\end{equation}
where, $y_t^* = \log\{ y_t / (1-y_t)\},$ $\mu_t^* = \psi(\mu_t\phi_t)-\psi((1-\mu_t)\phi_t)$ and $v_t$ is given in \ref{A3} in the Appendix.
Hence, we can write 
$\check{y}_t - \hat{\check{y}}_{(t)} = {r^{\beta}_t}/({1 - h_{tt}^*}),
$
where $h^*_{tt}$ is the $t$th diagonal element of projection matrix $$H^* =
 ({\,\widehat{\! W}}{\,\widehat{\! \Phi}})^{1/2}J_1(J_1{\,\widehat{\! \Phi}}{\,\widehat{\! W}}J_1)^{-1}J_1^{\top}({\,\widehat{\! \Phi}}{\,\widehat{\! W}})^{1/2}.$$
 Finally, for the nonlinear beta regressions models the classic PRESS statistic based on \eqref{press:1} is given by
\begin{equation}\label{eq4}
	PRESS = \sum^{n}_{t=1}(\check{y}_t - \hat{\check{y}}_{(t)})^2 = \sum^{n}_{t=1}\left(\frac{r^{\beta}_t}{1 - h_{tt}^*}\right)^2. 
\end{equation}

Note that  the  $t$th observation in \eqref{eq4} is not used in fitting the regression model to predict $y_t$, then both the external predicted values $\hat{y}_{(t)}$ and the external residuals $e_{(t)}$ are independent of $y_t$. This fact  enables the PRESS statistic to be a true assessment of the prediction capabilities of the regression model regardless of the overall model fit quality. Additionally, when the predictors in \eqref{component} are linear functions of the parameters, i.e., $g(\mu_t)=x^\top_t\beta$ and $ h(\phi_t)=z^\top_t\gamma,$ the expression in \eqref{eq4} also represent the PRESS statistic for a class of linear beta regression models with $p = k + q$ unknown
regression parameters.

Considering the same approach to construct the determination coefficient $R^2$ for linear models, we can think in a prediction coefficient based on PRESS, namely 
\begin{equation}\label{eq5}
P^2 = 1 - \frac{PRESS}{SST_{(t)}},
\end{equation}
\noindent where $SST_{(t)} = \sum_{t=1}^n(y_t- \bar{y}_{(t)})^2$ and  $\bar{y}_{(t)}$ is the arithmetic average of the ${y}_{(t)},\,t = 1,\ldots, n.$.  It can be shown that $SST_{(t)} = (n/(n-p))^2 SST$, wherein $p$ is the number of model parameters and $SST$ is the Total Sum of Squares for the full data. For a class of beta regressions models with varying dispersion,  
$SST = \sum_{t=1}^n(\check{y}_t- \bar{\check{y}})^2$, $\bar{\check{y}}$ is the  is the arithmetic average of the  $\check{y}_t = \widehat{\phi}_t^{1/2}\widehat{w}_t^{1/2} u_{1,t},\,t = 1,\ldots, n.$
% and $p = k + q$.
It is noteworthy that the measures $R^2$ and $P^2$ are distinct, since that the $R^2$ propose to measure the model fit quality  and the $P^2$ measure the predictive power of the model. 

\cite{CookWeisberg82book} suggest other versions of PRESS statistic based on different choices of residuals. Thus, we present another version of PRESS statistic and $P^2$ measure by considering the combined residual proposed by \cite{ESPINHEIRA2017}. In this way, 
\begin{equation}\label{eq6}
PRESS_{\beta\gamma}  = \sum^{n}_{t=1}\left(\frac{r^{\beta\gamma }_{p,t}}{1 - h_{tt}^*}\right)^2\quad {\rm and} \quad
P^2_{\beta\gamma} = 1 - \frac{PRESS_{\beta\gamma}}{SST_{(t)}}, 
\end{equation}
\noindent respectively, where  
%In (\ref{eq6}) $r^{\beta\gamma }_{p,t}$ is the combined residual proposed by \cite{ESPINHEIRA2017} and defined as
\begin{equation}
\begin{split}\label{eq7}
&r^{\beta\gamma}_{t} = \frac{(y^{*}_t - {\,\widehat{\! \mu}}^{*}_t) +{\,\widehat{\! a}}_t}{\sqrt{\,\widehat{\! \zeta_t}}}, \quad
a_t = \mu_t(y_t^{*} - \mu_{t}^{*}) + \log(1 - y_t) - \psi((1 - \mu_t)\phi_t) + \psi(\phi_t)\\&\mbox{and}\quad \zeta_t = (1  + \mu_t)^2  \psi^{\prime}(\mu_t\phi_t) + \mu_t^2
\psi^{\prime}((1 - \mu_t)\phi_t) - \psi^{\prime}(\phi_t).
\end{split}
\end{equation}
%
%
%
%
%It is noteworthy that the measures $R^2$ and $P^2$ are distinct, since that the $R^2$ propose to measure the model fit quality  and the $P^2$ and $P^2_{\beta\gamma}$  measure the predictive power of the model. 
%Note that, $P^2$ and $P^2_{\beta\gamma}$ given in \eqref{eq5} and  \eqref{eq6}, respectively, are not positive quantifiers. In fact, the $\text {PRESS/SST}_{(t)}$  is  a positive quantity, thus the  $P^2$ and the $P^2_{\beta\gamma}$  take values in $(-\infty; 1]$. The closer to one the better is the model predictive power.
%
Note that, $P^2$ and $P^2_{\beta\gamma}$ given in \eqref{eq5} and \eqref{eq6}, respectively, are not positive quantifiers. In fact, the $\text {PRESS/SST} _ {(t)} $ is a positive quantity, thus the $P^2$ and the $P^2_{\beta\gamma}$ are measures that take values in $(-\infty; 1] $. The closer to one the better is the predictive power of the model.

In order to check the model goodness-of-fit with linear or nonlinear predictors for a class of beta regression, we evaluate the $R^2_{FC}$ defined as the square of the sample coefficient of correlation between $g(y)$ and $\widehat \eta_1$ \citep{Ferrari2004}, and its penalized version based on \cite{BAYER2017} given by
$
R^2_{FC_{c}} = 1 - (1 - R^2_{FC})(n-1)/(n-(k_1+q_1)), 
$
where $k_1$ and $q_1$ are, respectively, the number of covariates of the mean submodel and dispersion submodel.

We also evaluate two version of pseudo-$R^2$ based on likelihood ratio. The first one proposed by \cite{Nagelkerke1991}:
$
R^2_{LR} = 1 - ({L_{null}}/{L_{fit}})^{2/n} 
$,
where $L_{null}$ is the maximum likelihood achievable (saturated model) and $L_{fit}$ is the achieved by the model under
investigation.
The second one is a proposal of \cite{BAYER2017} that takes account the inclusion of covariates both in the mean submodel and  in the precision submodel,  given by:
$$
R^2_{{LR}_c} =1 - (1 - R^2_{LR})\left(\frac{n-1}{n-(1+\alpha)k_1-(1-\alpha)q_1}\right)^\delta, 
$$
where $\alpha\in[0,1]$ and $\delta>0$.  Based on simulation results obtained in
 \cite{BAYER2017}  we choose in this work the values $\alpha=0.4$ and $\delta = 1$.
%authors indicate the choice 
%
%
%Following the results present by the authors we choose $\alpha=0.4$ and $\delta = 1$.
%
%
%In this work and the analog form of the \cite{BAYER2017}
%Similarly we 
%and define 
Therefore, penalized versions of $P^2$ and $P^2_{\beta\gamma}$ are respectively now given by:
$
P^2_c =1 - (1 - P^2)(n-1)/(n-(k_1+q_1)) 
$
and
$
P^2_{{\beta\gamma}_c} =1 - (1 - P^2_{\beta\gamma})(n-1)/(n-(k_1+q_1)).
$

\section{Simulation}

In this section we
simulate several different data generating processes to evaluate the performance
of the predictive measures. The Monte Carlo experiments were carried out  using both fixed and varying dispersion beta regressions as data generating processes. All results are based on 10,000 Monte Carlo replications. 

\paragraph{Linear models:} 
%\subsection{Linear models}
%The Monte Carlo experiments were carried out  using both fixed and varying dispersion beta regressions as data generating processes. All results are based on 10,000 Monte Carlo replications.  
Table~\ref{T:T1} shows the  mean values of the predictive statistics obtained by simulation  of fixed dispersion beta regression model that involves a systematic component for the mean given by
%Table~\ref{T:T1} shows the  mean values of the statistics obtained by simulation  relative to the specification of fixed dispersion beta regression model that involves a systematic component for the mean given by
%%
%
%
%
%
%as data generating processe, given
%by the systematic 
%
% shows the simulation results relative to the specification
%
%
%
\begin{equation}
\label{mod:01}
\log \left(\frac{\mu_t}{1-\mu_t}\right) = \beta_1 + \beta_2\,x_{t2} + \beta_3\,x_{t3} + \beta_4\,x_{t4} + \beta_5\,x_{t5}, \quad t = 1, \ldots, n ,
\end{equation}
The covariate values were independently obtained as
random draws of the following distributions: $X_{ti} \sim U(0,1)$, $i = 2, \ldots, 5$  and
were kept fixed throughout the experiment.
 The precisions, the sample sizes and  the range of mean  response are, respectively,  $\phi = (20, 50, 148, 400)$, $n = (40, 80, 120, 400)$, $\mu \in (0.005, 0.12)$, $\mu \in (0.90, 0.99)$ and $\mu\in (0.20, 0.88)$.
 Under the  model specification given in \eqref{mod:01} we investigate the performances of the statistics by omitting covariates. In this case, %under omission of covariates. Here, 
 we considered the Scenarios 1, 2 and 3, in which are omitted, three, two and one covariate, respectively. In a fourth scenario the estimated model is correctly specified (true model).
%  In Table~\ref{T:T1} we present the mean values of the statistics  from 10,000 Monte Carlo replications, for some values of $\phi$ and $n$. In this table we only considere  the versions of the $P^2$ and $R^2_{LR}$ statistics.

The results in Table~\ref{T:T1} show that the mean values of all statistics increase as covariates are included in the model and the value of $\phi$ increases.
On the other hand, as the size of the sample increases, the misspecification of the model is evidenced by lower values of the statistics (Scenarios 1, 2 and 3). It shall be noted that the values of all statistics  are considerably larger when $\mu \in (0.20,0.88)$. Additionally, its values approaching one when the estimated model is closest to the true model. For instance, in Scenario 4 for $n = 40$, $\phi = 150$ the values of $P^{2}$  and  $R^2_{LR}$  are, respectively,  $0.936$  and  $0.947$. 

The behavior of the statistics   for  finite samples  changes  substantially  when $\mu \in (0.90; 0.99)$. It is  noteworthy 
the reduction of its values, revealing the difficulty  in fitting the model and make prediction when  $\mu \approx 1$ (The log-likelihood of the model tends to no longer limited).
Indeed, in this range of $\mu$ is more difficult to make prediction than to fit the model. For example, in Scenario 1, when three covariates are omitted from the model,
$n = 40$  and $\phi = 150$  the $P^{2}$ value equals to 0.071,  whereas the $R^2_{LR}$ value is 0.243. Similar results were obtained for   $n = 80, 120$. 
Even under true specification (Scenario 4) the model predictive power is more affected than the model quality of fit  by the fact of $\mu \approx 1$.
For instance, when  $n=120$  and  $\phi = 50$  we have $P^{2}_{\beta\gamma} = 0.046$ and
$R^2_{LR} = 0.565$. 
The same difficulty in obtaining predictions and in fitting the regression model  occurs when $\mu \in (0.005, 0.12)$. Once again the greatest difficulty   lies on the predictive sense.

\begin{table}[htp!]
\begin{center}
	\caption{\label{T:T1} Values of the statistics. True model versus
%	
%	:
%		$g(\mu_t) =\log( {\mu_t}/{(1-\mu_t)}) = \beta_1 + \beta_2\,x_{t2} + \beta_3\,x_{t3} + \beta_4\,x_{t4} + \beta_5\,x_{t5}$ and $\phi$ constant across observations.
%%		$x_{ti} \sim U(0,1)$,  $\,\,i=2,3,4,5,$ $\,t=1,\ldots,n$ and $\phi$ constant across observations.
		misspecification models (omitted covariates (Scenarios 1, 2 and 3)).}
		\medskip
	\renewcommand{\tabcolsep}{0.48pc} % enlarge column spacing
	\renewcommand{\arraystretch}{1.31}
	%\smallskip
	%\centering
	%\fbox{%
	%\begin{tabular}{@{}|c|c|c|c|c|c|c|c|c|c|c|c|c|c|} \hline
	{\tiny
		\begin{tabular}{*{14}{c}} \hline
			&
%			\textbf{Scenarios}
			&\multicolumn{3}{c}{
			\textbf{Scenario 1}}&\multicolumn{3}{c}{
			\textbf{Scenario 2}}&\multicolumn{3}{c}{
			\textbf{Scenario 3}}&\multicolumn{3}{c}{
			\textbf{Scenario 4}}\\\cmidrule(lr){3-5}   \cmidrule(lr){6-8}    \cmidrule(lr){9-11}  \cmidrule(lr){12-14}
			&Estimated   &\multicolumn{3}{c}{$g(\mu_t) = \beta_1 + \beta_2\,x_{t2}$}&\multicolumn{3}{c}{$g(\mu_t) = \beta_1 + \beta_2\,x_{t2}$}&\multicolumn{3}{c}{$g(\mu_t) = \beta_1 + \beta_2\,x_{t2}$}&\multicolumn{3}{c}{$\hskip-0.1trueing(\mu_t) = \beta_1 + \beta_2\,x_{t2}+\hskip-0.1truein$}\\
			&model&\multicolumn{3}{c}{}&\multicolumn{3}{c}{$+ \beta_3\,x_{t3}$}&\multicolumn{3}{c}{$+ \beta_3\,x_{t3} + \beta_4\,x_{t4}$}&\multicolumn{3}{c}{$\hskip-0.05truein\beta_3\,x_{t3} + \beta_4\,x_{t4}+\beta_5\,x_{t5}\hskip-0.1truein$}\\ \cmidrule(lr){3-14}
		& \multicolumn{13}{c}{$\mu \in(0.20, 0.88);\quad\quad \beta=(-1.9,1.2,1.0,1.1,1.3)^{\!_\top}$.} \\ \cmidrule(lr){2-14}
			$n$&$\phi \rightarrow $&20&50&150&20&50&150&20&50&150&20&50&150\\ \cmidrule(lr){2-14}
			\multirow{ 6}{*}{40} &$P^2$                      &0.307&0.363&0.393&0.393&0.463&0.502&0.506&0.602&0.656&0.694&0.847&0.936\\
			&$P^2_c$                    &0.270&0.329&0.361&0.342&0.418&0.461&0.450&0.557&0.617&0.649&0.825&0.927\\
			&$P^2_{\beta\gamma}$        &0.304&0.357&0.385&0.402&0.473&0.512&0.512&0.609&0.663&0.694&0.847&0.936\\
			&$P^2_{{\beta\gamma}_c}$    &0.267&0.322&0.352&0.352&0.429&0.472&0.456&0.564&0.625&0.649&0.825&0.927\\
			&$R^2_{LR}$                 &0.296&0.358&0.391&0.394&0.473&0.515&0.518&0.620&0.675&0.723&0.869&0.947\\
			&$R^2_{{LR}_c}$             &0.258&0.324&0.358&0.344&0.429&0.475&0.463&0.577&0.638&0.682&0.849&0.939\\\hline 
			\multirow{ 6}{*}{80} &$P^2$                      &0.286&0.346&0.379&0.368&0.445&0.488&0.506&0.584&0.643&0.666&0.833&0.930\\     
			&$P^2_c$                    &0.267&0.329&0.363&0.343&0.423&0.468&0.450&0.561&0.624&0.643&0.821&0.925\\     
			&$P^2_{\beta\gamma}$        &0.282&0.339&0.370&0.377&0.455&0.498&0.512&0.590&0.650&0.666&0.833&0.930\\     
			&$P^2_{{\beta\gamma}_c}$    &0.264&0.322&0.353&0.353&0.433&0.478&0.456&0.569&0.632&0.643&0.821&0.925\\     
			&$R^2_{LR}$                 &0.291&0.356&0.391&0.385&0.468&0.513&0.518&0.614&0.672&0.706&0.860&0.943\\     
			&$R^2_{{LR}_c}$             &0.273&0.339&0.375&0.361&0.447&0.494&0.463&0.593&0.655&0.686&0.851&0.939\\\hline
			\multirow{ 6}{*}{120}&$P^2$                      &0.279&0.340&0.374&0.360&0.439&0.483&0.469&0.578&0.639&0.656&0.833&0.928\\     
			&$P^2_c$                    &0.267&0.329&0.363&0.343&0.424&0.470&0.450&0.563&0.626&0.641&0.821&0.925\\     
			&$P^2_{\beta\gamma}$        &0.275&0.333&0.365&0.369&0.449&0.493&0.475&0.585&0.646&0.656&0.833&0.928\\     
			&$P^2_{{\beta\gamma}_c}$    &0.263&0.322&0.354&0.353&0.435&0.480&0.457&0.570&0.634&0.641&0.821&0.925\\     
			&$R^2_{LR}$                 &0.290&0.355&0.390&0.382&0.467&0.513&0.501&0.612&0.671&0.700&0.860&0.942\\     
			&$R^2_{{LR}_c}$             &0.278&0.344&0.380&0.366&0.453&0.500&0.483&0.598&0.660&0.686&0.851&0.939\\ \hline %\cmidrule(lr){2-14}
		& \multicolumn{13}{c}{$\mu \in (0.90, 0.99);\quad\quad \beta=(1.8,1.2,1,1.1,0.9)^{\!_\top}$.}\\ \cmidrule(lr){2-14}
			$n$&$\phi \rightarrow $&20&50&150&20&50&150&20&50&150&20&50&150\\ \cmidrule(lr){2-14}
			\multirow{ 6}{*}{40} &$P^2$                      &0.119&0.061&0.071&0.139& 0.062& 0.072&0.171& 0.072&0.156&0.149& 0.089&0.213\\       
			&$P^2_c$                    &0.071&0.010&0.021&0.067&$-0.016$&$-0.006$&0.076&$-0.034$&0.059&0.023&$-0.045$&0.097\\       
			&$P^2_{\beta\gamma}$        &0.119&0.061&0.071&0.139& 0.062& 0.071&0.171& 0.072&0.155&0.148& 0.089&0.213\\       
			&$P^2_{{\beta\gamma}_c}$    &0.072&0.010&0.020&0.068&$-0.016$&$-0.007$&0.076&$-0.034$&0.058&0.022&$-0.045$&0.097\\       
			&$R^2_{LR}$                 &0.164&0.196&0.243&0.221& 0.266& 0.336&0.271& 0.374&0.466&0.444& 0.593&0.774\\       
			&$R^2_{{LR}_c}$             &0.119&0.153&0.203&0.157& 0.205& 0.281&0.188& 0.303&0.405&0.362& 0.533&0.741\\\hline 
			\multirow{ 6}{*}{80} &$P^2$                      &0.093&0.036&0.044&0.112& 0.038& 0.046&0.149& 0.045&0.120&0.123& 0.056&0.175\\       
			&$P^2_c$                    &0.070&0.011&0.019&0.077& 0.000& 0.008&0.103&$-0.006$&0.073&0.063&$-0.007$&0.119\\       
			&$P^2_{\beta\gamma}$        &0.094&0.036&0.043&0.113& 0.038& 0.045&0.149& 0.045&0.119&0.122& 0.056&0.175\\       
			&$P^2_{{\beta\gamma}_c}$    &0.070&0.011&0.018&0.078& 0.000& 0.008&0.104&$-0.006$&0.072&0.063&$-0.008$&0.119\\       
			&$R^2_{LR}$                 &0.158&0.190&0.240&0.211& 0.253& 0.327&0.268& 0.356&0.451&0.416& 0.571&0.760\\       
			&$R^2_{{LR}_c}$             &0.136&0.169&0.221&0.180& 0.224& 0.301&0.229& 0.321&0.422&0.376& 0.542&0.744\\\hline 
			\multirow{ 6}{*}{120}&$P^2$                      &0.085&0.028&0.035&0.102& 0.030& 0.038&0.141& 0.036&0.107&0.114&0.046 &0.162\\       
			&$P^2_c$                    &0.069&0.012&0.018&0.079& 0.005& 0.013&0.111& 0.002&0.076&0.075&0.004 &0.125\\       
			&$P^2_{\beta\gamma}$        &0.085&0.028&0.034&0.103& 0.030& 0.037&0.141& 0.036&0.107&0.113&0.046 &0.162\\       
			&$P^2_{{\beta\gamma}_c}$    &0.069&0.012&0.018&0.080& 0.005& 0.012&0.112& 0.002&0.076&0.075&0.004 &0.125\\       
			&$R^2_{LR}$                 &0.156&0.188&0.239&0.207& 0.249& 0.324&0.268& 0.349&0.447&0.406&0.565 &0.756\\       
			&$R^2_{{LR}_c}$             &0.142&0.175&0.226&0.186& 0.230& 0.306&0.242& 0.327&0.428&0.380&0.545 &0.745\\ \hline %\cmidrule(lr){2-14} 
		      & \multicolumn{13}{c}{$\mu \in(0.005, 0.12);\quad\quad \beta=(-1.5,-1.2,-1.0,-1.1,-1.3)^{\!_\top}$.}\\ \cmidrule(lr){2-14}
			$n$&$\phi \rightarrow $&20&50&150&20&50&150&20&50&150&20&50&150\\ \cmidrule(lr){2-14} 
			\multirow{ 6}{*}{40} &$P^2$                      &0.128&0.063&0.056&0.108& 0.059& 0.028&0.153& 0.070&0.202&0.149& 0.090&0.212\\       
			&$P^2_c$                    &0.081&0.013&0.005&0.033&$-0.020$&$-0.053$&0.056&$-0.036$&0.111&0.023&$-0.044$&0.096\\       
			&$P^2_{\beta\gamma}$        &0.128&0.063&0.057&0.107& 0.055& 0.026&0.153& 0.071&0.203&0.150& 0.090&0.212\\       
			&$P^2_{{\beta\gamma}_c}$    &0.081&0.013&0.006&0.032&$-0.023$&$-0.055$&0.056&$-0.036$&0.112&0.025&$-0.044$&0.097\\       
			&$R^2_{LR}$                 &0.199&0.215&0.254&0.265& 0.349& 0.379&0.326& 0.415&0.548&0.442& 0.595&0.774\\       
			&$R^2_{{LR}_c}$             &0.156&0.172&0.214&0.204& 0.295& 0.327&0.249& 0.348&0.496&0.360& 0.535&0.741\\\hline 
			\multirow{ 6}{*}{80} &$P^2$                      &0.105&0.040&0.032&0.083& 0.043& 0.012&0.128& 0.038&0.165&0.123& 0.057&0.174\\       
			&$P^2_c$                    &0.081&0.015&0.006&0.047& 0.005&$-0.027$&0.081&$-0.013$&0.121&0.064&$-0.007$&0.119\\       
			&$P^2_{\beta\gamma}$        &0.104&0.040&0.032&0.081& 0.039& 0.010&0.128&0.038&0.166&0.124&0.057&0.175\\       
			&$P^2_{{\beta\gamma}_c}$    &0.081&0.015&0.007&0.045& 0.001&$-0.030$&0.081&$-0.013$&0.121&0.065&$-0.007$&0.119\\       
			&$R^2_{LR}$                 &0.197&0.211&0.251&0.253& 0.340& 0.372&0.311& 0.394&0.534&0.416& 0.572&0.760\\       
			&$R^2_{{LR}_c}$             &0.176&0.191&0.231&0.223& 0.314& 0.347&0.274& 0.362&0.509&0.376& 0.543&0.743\\\hline 
			\multirow{ 6}{*}{120}&$P^2$                      &0.097&0.033&0.024&0.074& 0.037& 0.006&0.118& 0.028&0.153&0.114& 0.046&0.162\\       
			&$P^2_c$                    &0.081&0.016&0.007&0.050& 0.012&$-0.020$&0.088&-0.006&0.123&0.075& 0.004&0.125\\       
			&$P^2_{\beta\gamma}$        &0.096&0.032&0.024&0.072& 0.034& 0.004&0.118& 0.028&0.153&0.115& 0.046&0.162\\       
			&$P^2_{{\beta\gamma}_c}$    &0.081&0.016&0.007&0.048& 0.009&$-0.022$&0.087&$-0.006$&0.124&0.076& 0.004&0.125\\       
			&$R^2_{LR}$                 &0.195&0.209&0.250&0.247& 0.337& 0.370&0.304& 0.388&0.530&0.407& 0.565&0.755\\       
			&$R^2_{{LR}_c}$             &0.181&0.196&0.237&0.228& 0.320& 0.354&0.280& 0.367&0.513&0.381& 0.546&0.744\\\hline 
		\end{tabular}}
		%}
		\end{center}
	\end{table}

%In it follows we shall investigate the empirical distributions of the statistics. To that end we built the boxplots of the 10,000  values of the statistics obtained from the Monte Carlo simulations.  
%We must emphasize that in all boxplots the ``side point" represents the mean value of the  replications of the  statistics. We also built boxplots to evaluate the performance of $R^2_{{FC}}$ and its penalized version. 

\begin{figure}[!htp]
	\centering
	\vskip-0.8truein
	\makebox{\includegraphics[width=4.0truein,angle=270]{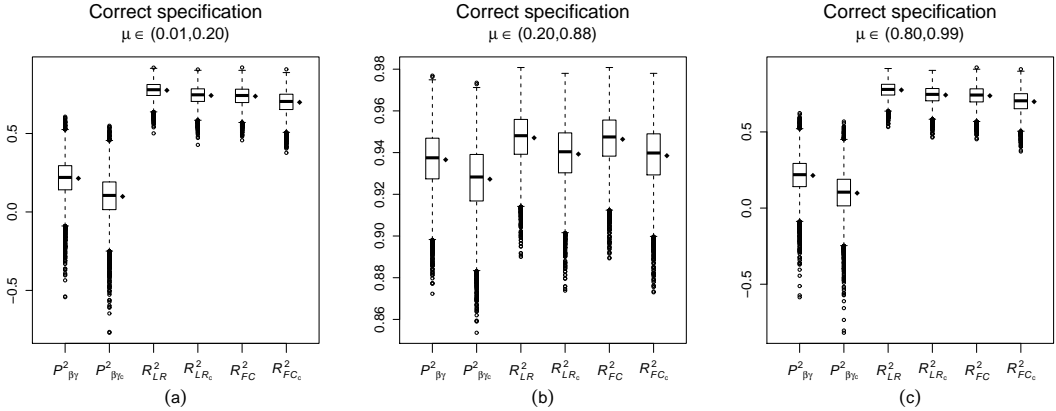}}
	\vskip-1.0truein
	\caption{\label{fig:1} Model estimated correctly: $g(\mu_t) = \beta_1 + \beta_2\,x_{t2} +\beta_3\,x_{t3}+ \beta_4\,x_{t4}+ \beta_5\,x_{t5}$. $\mu \in(0.20, 0.88); \quad \beta=(-1.9,1.2,1.0,1.1,1.3)^\top$; $\mu \in (0.90, 0.99);\quad\beta=(1.8,1.2,1,1.1,0.9)^\top$; $\mu \in(0.005, 0.12);\quad\beta=(-1.5,-1.2,-1.0,-1.1,-1.3)^\top$.}
%	\vskip-0.4truein
\end{figure}

 Figure~\ref{fig:1} present the boxplots of the 10,000 replications of the statistics:  $P^2_{{\beta\gamma}}$, $P^2_{{\beta\gamma}_c}$, $R^2_{LR}$, $R^2_{{LR}_c}$, $R^2_{FC}$ and $R^2_{{FC}_c}$  when the model is correctly specified (scenario 4), $n=40$ and $\phi=150$. In all boxplots the ``side point" represents the mean value of the replications of the  statistics.  In the panel (a) we present the boxplots when $\mu \approx 0$. In the panel (b) we present the boxplots when $\mu$ is scattered on the standard unit interval and in the panel (c) we present the boxplots for $\mu \approx 0$. This figure  shows that the means and  the medians of all statistics are close. We also can notice based on the Figure~\ref{fig:1} that  both prediction power  and goodness-of-fit of the model  are affected when $\mu$ is close to the boundaries of the standard unit interval.  
However, it is noteworthy the great difficult to make prediction. 
Additionally, is possible to notice that the versions of $R^2$ displays similar behavior. In it follows we shall investigate the empirical distributions behaviour of the statistics proposed.
 
\begin{figure}[!htp]
	\centering
%	\vskip-1.0truein
	\makebox{\includegraphics[width=4.0truein,angle=270]{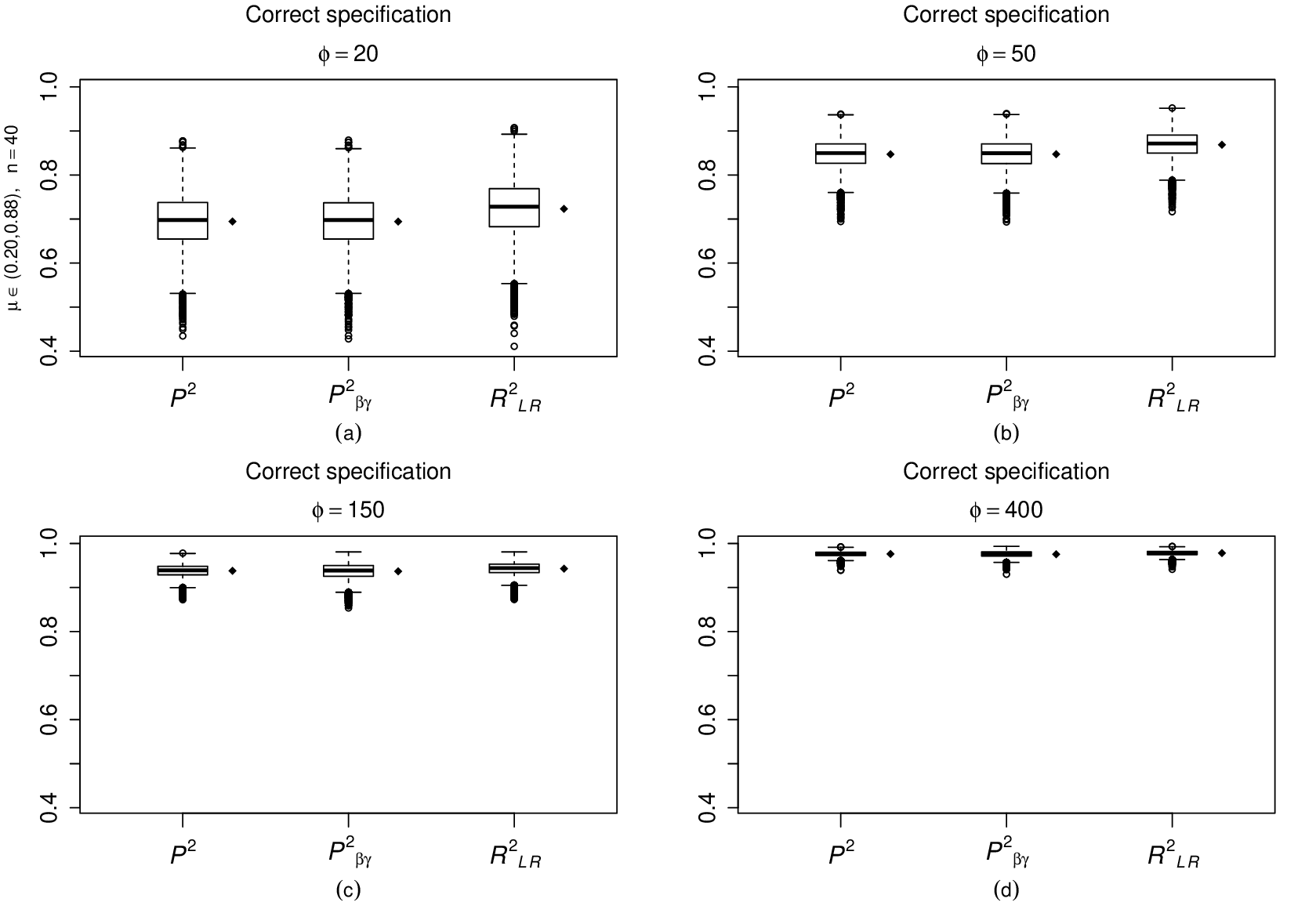}}
%	\vskip-1.2truein
	\caption{\label{fig:2} Model estimated correctly: $g(\mu_t) = \beta_1 + \beta_2\,x_{t2} +\beta_3\,x_{t3}+ \beta_4\,x_{t4}+ \beta_5\,x_{t5}$. $\mu \in(0.20, 0.88);\quad \beta=(-1.9,1.2,1.0,1.1,1.3)^\top$. 
	}
\end{figure}
\begin{figure}[!htp]
	\centering
%	\vskip-0.9truein
	\makebox{\includegraphics[width=4.0truein,angle=270]{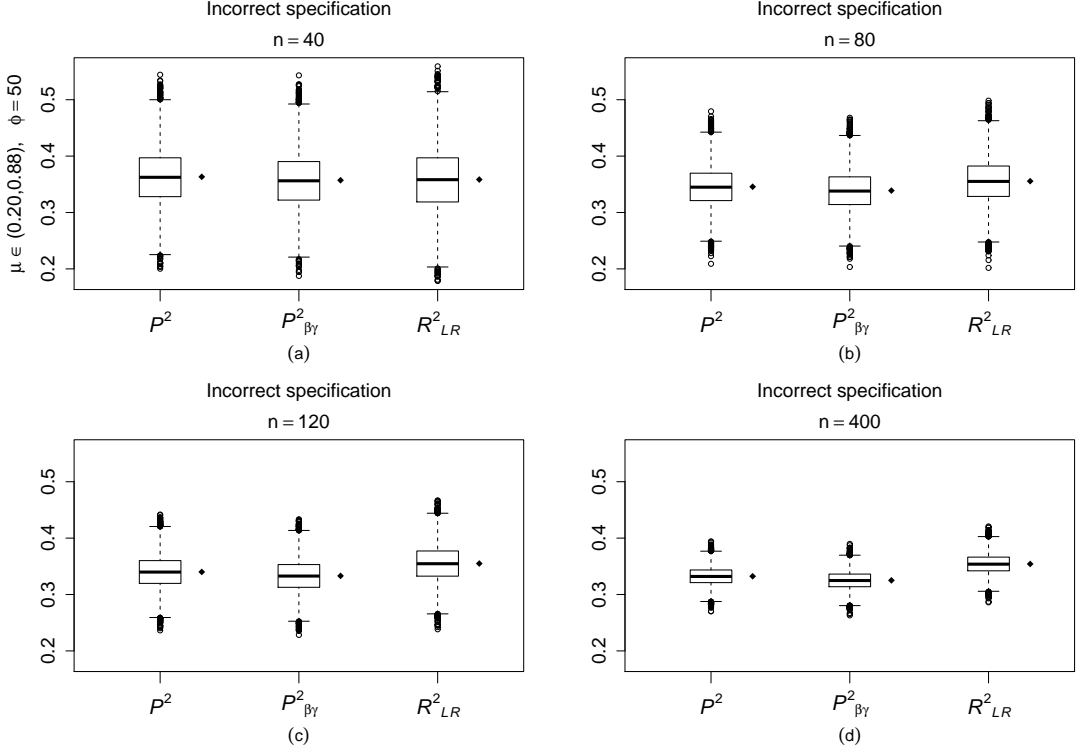}}
%	\vskip-1.0truein
	\caption{\label{fig:3} Omitted covariates. Estimated model: $g(\mu_t) = \beta_1 + \beta_2\,x_{t2}$. Correct model: $g(\mu_t) = \beta_1 + \beta_2\,x_{t2} +\beta_3\,x_{t3}+ \beta_4\,x_{t4}+ \beta_5\,x_{t5}$; $\mu \in(0.20, 0.88);\quad \beta=(-1.9,1.2,1.0,1.1,1.3)^\top$ }
\end{figure}

In Figures~\ref{fig:2} and ~\ref{fig:3} we consider $\mu\in(0.20,0.88)$. In Figure~\ref{fig:2} the model is estimated correctly,  $n = 40$ and $\phi = (20, 50, 150, 400)$.  We notice that the prediction power distortions increase as the precision parameter increases, as expected.  In Figure~\ref{fig:3} we consider a misspecification problem (three omitted covariates). For illustration, we consider only $\phi = 50$ and $n=40,80,120,400$. It is important to notice that here the performance of the prediction measures  does not deteriorate when the sample size is increased.  Based on these figures we can reach some conclusions. First, the model precision affects both its predict power and goodness-of-fit. Second, for this range of $\mu$ the performance of the statistics are similar revealing the correct specification of the model (Figure~\ref{fig:2}). Third, when three covariates are omitted, with the increasing of sample size the replications values of the statistics tend  being concentrated at small values due to the misspecification problem (Figure~\ref{fig:3}).

In what follows, we shall report simulation results on the finite-sample performance of the statistics  when the  dispersion modeling is neglected. To that end, the true data generating process considers  varying dispersion, but a fixed dispersion beta regression is estimated. We also used different covariates in the mean and precision submodels. The samples sizes are $n = 40, 80, 120$. We generated 20 values for each covariate and replicated them to get covariate values for the three sample sizes (once, twice and three times, respectively). 

This was done so that the intensity degree of nonconstant dispersion
\begin{equation}\label{eq16}
\lambda =  \frac{\phi_{\max}}{\phi_{\min}}={\max\limits_{t=1,\ldots,n}\{\phi_t \}\over\min\limits_{t=1,\ldots,n}\{\phi_t\}},
\end{equation}
would remain constant as the sample size changes.  The numerical results were obtained using the following beta regression model
%True models:
$g(\mu_t) =\log( {\mu_t}/{(1-\mu_t)}) = \beta_1 + \beta_i\,x_{ti},\,{\text{and}}\,\log(\phi_t) = \gamma_1 + \gamma_i\,z_{ti},$ $x_{ti} \sim U(0,1)$, $z_{ti} \sim U(-0.5,0.5)$, $\,\,i=2,3,4,5,$ and $\,t=1,\ldots,n$
under different choices of parameters (Scenarios): 
 {\bf Scenario 5}: $\beta=(-1.3,3.2)^\top$, $\mu\in(0.22,0.87)$, $[\gamma=(3.5,3.0)^\top;\lambda\approx20],$ $[\gamma=(3.5,4.0)^\top;\lambda\approx 50]$ and  $[\gamma=(3.5,5.0)^\top;\lambda\approx150].$ 
 {\bf Scenario 6}: $\beta=(-1.9,1.2,1.6,2.0)^\top$, $\mu\in(0.24,0.88)$,	$[\gamma=(2.4,1.2,-1.7,1.0)^\top; \lambda\approx20]$, $[\gamma=(2.9,2.0,-1.7,2.0)^\top; \lambda\approx 50]$ and $[\gamma=(2.9,2.0,-1.7,2.8)^\top; \lambda\approx100]$.
{\bf Scenarios 7 {\rm and} 8 (Full models)}: $\beta=(-1.9,1.2,1.0,1.1,1.3)^\top$, $\mu\in(0.20,0.88)$, $[\gamma=(3.2,2.5,-1.1,1.9,2.2)^\top;  \lambda\approx20]$, $[\gamma=(3.2,2.5,-1.1,1.9,3.2)^\top; \lambda\approx50]$, and $[\gamma=(3.2,2.5,1.1,1.9,4.0)^\top; \lambda\approx100]$. All results were obtained using 10,000 replics Monte Carlo replications.

%Our aim is  to assess the performance of the statistics  when the  dispersion modeling is  neglected.  To that end, the true data generating process considers  varying dispersion,   
%but a fixed dispersion beta regression is estimated; see
%Table~\ref{T:T2} (Scenarios 5, 6 and 7). We also displayed in Table~\ref{T:T2}  the  values of the statistics when the model is correctly specified (Scenario 8). True models:
%$g(\mu_t) =\log( {\mu_t}/{(1-\mu_t)}) = \beta_1 + \beta_i\,x_{ti},\,{\text{and}}\,\log(\phi_t) = \gamma_1 + \gamma_i\,z_{ti},$ $x_{ti} \sim U(0,1)$, $z_{ti} \sim U(-0.5,0.5)$, $\,\,i=2,3,4,5,$ and $\,t=1,\ldots,n.$
% {\bf Scenario 5}: $\beta=(-1.3,3.2)^\top$, $\mu\in(0.22,0.87)$, $\gamma=(3.5,3.0)^\top;\lambda\approx20, \gamma=(3.5,4.0)^\top;\lambda\approx 50, \, \gamma=(3.5,5.0)^\top;\lambda\approx150$.
%{\bf Scenario 6}: $\beta=(-1.9,1.2,1.6,2.0)^\top$, $\mu\in(0.24,0.88)$,	$\gamma=(2.4,1.2,-1.7,1.0)^\top$; $\lambda\approx20$, $\gamma=(2.9,2.0,-1.7,2.0)^\top$; 
%$\lambda\approx 50$ and $\gamma=(2.9,2.0,-1.7,2.8)^\top$; $\lambda\approx100$.
%{\bf Scenarios 7 and 8}: $\beta=(-1.9,1.2,1.0,1.1,1.3)^\top$, $\mu\in(0.20,0.88)$, $\gamma=(3.2,2.5,-1.1,1.9,2.2)^\top$; $ \lambda\approx20$, $\gamma=(3.2,2.5,-1.1,1.9,3.2)^\top$; $\lambda\approx50$, and $\gamma=(3.2,2.5,1.1,1.9,4.0)^\top$; $\lambda\approx100$.

Table~\ref{T:T2}  contain the  values of the statistics. We notice that for each scenario the prediction power measure not present  high distortion  when we increase intensity degree of nonconstant dispersion. However, in the case of misspecification the statistics display smaller values in comparison with Scenario 8 (True specification), in which as greater is $\lambda$ greater are the values of the statistics, as expected.  Other important impression lies in the fact that the values of the $R^2_{FC}$ are considerably smaller than the values of  the others statistics, in special when $\lambda$ increases.

 That is a strong evidence that the $R^2 _{FC}$  does not perform well under nonconstant dispersion models. In fact, under nonconstant dispersion models the better performances are of the $P^2$ statistics, both in identifying wrong and correct specifications.

\begin{table}[htp!]
\begin{center}
	\caption{\label{T:T2} Values of the statistics. Misspecified models, $\phi$ fixed: Scenarios 5, 6 and 7 versus Scenario 8 (correct specification).} 
	\medskip
	\renewcommand{\tabcolsep}{0.65pc} % enlarge column spacing
	\renewcommand\arraystretch{1.53}% enlarge row spacing                     
	%\centering
	%\fbox{%
	{\tiny
		\begin{tabular}{*{14}{c}}\hline 
		% {\scriptsize
		% \begin{tabular}{@{}c|c|c|c|c|c|c|c|c|c|c|c|c|c} \hline
%		\text{Scenarios}
		&\multicolumn{3}{c}{\textbf{Scenario 5}}&\multicolumn{3}{c}{\textbf{Scenario 6}}&\multicolumn{3}{c}{\textbf{Scenario 7}}&\multicolumn{3}{c}{\textbf {Scenario 8}}\\\cmidrule(lr){2-4}   \cmidrule(lr){5-7}    \cmidrule(lr){8-10}    \cmidrule(lr){11-13}
		&\multicolumn{3}{c}{$g(\mu_t) = \beta_1 + \beta_2\,x_{t2}$}    &\multicolumn{3}{c}{$g(\mu_t) = \beta_1 + \beta_2\,x_{t2} +$}      &\multicolumn{3}{c}{$g(\mu_t) = \beta_1 + \beta_2\,x_{t2} +$}                   &\multicolumn{3}{c}{$g(\mu_t) = \beta_1 + \beta_2\,x_{t2} +$}               \\ 
		True            &\multicolumn{3}{c}{$+ \beta_3\,x_{t3}$}                       &\multicolumn{3}{c}{$\beta_3\,x_{t3} + \beta_4\,x_{t4}$}            &\multicolumn{3}{c}{$\beta_3\,x_{t3} + \beta_4\,x_{t4} + \beta_5\,x_{t5}$}      &\multicolumn{3}{c}{$\beta_3\,x_{t3} + \beta_4\,x_{t4} + \beta_5\,x_{t5}$}   \\                                                             
		models          &\multicolumn{3}{c}{$h(\phi_t) = \gamma_1 + \gamma_2\,z_{t2}$} &\multicolumn{3}{c}{$h(\phi_t) = \gamma_1 + \gamma_2\,z_{t2}+$}    &\multicolumn{3}{c}{$h(\phi_t) = \gamma_1 + \gamma_2\,z_{t2}+$}                 &\multicolumn{3}{c}{$h(\phi_t) = \gamma_1 + \gamma_2\,z_{t2}+$}               \\ 
		&\multicolumn{3}{c}{$+ \gamma_3\,z_{t3}$}                      &\multicolumn{3}{c}{$+ \gamma_3\,z_{t3} + \gamma_4\,z_{t4}$}        &\multicolumn{3}{c}{$\gamma_3\,z_{t3} + \gamma_4\,z_{t4} + \gamma_5\,z_{t5} $}  &\multicolumn{3}{c}{$\gamma_3\,z_{t3} + \gamma_4\,z_{t4} + \gamma_5\,z_{t5} $}\\ \cmidrule(lr){2-4}   \cmidrule(lr){5-7}    \cmidrule(lr){8-10}    \cmidrule(lr){11-13}                                                    
		       &\multicolumn{3}{c}{$g(\mu_t) = \beta_1 + \beta_2\,x_{t2}$}    &\multicolumn{3}{c}{$g(\mu_t) = \beta_1 + \beta_2\,x_{t2} +$}      &\multicolumn{3}{c}{$g(\mu_t) = \beta_1 + \beta_2\,x_{t2} +$}                   &\multicolumn{3}{c}{$g(\mu_t) = \beta_1 + \beta_2\,x_{t2} +$}                  \\ 
		Estimated          &\multicolumn{3}{c}{$+ \beta_3\,x_{t3}$}                       &\multicolumn{3}{c}{$+ \beta_3\,x_{t3} + \beta_4\,x_{t4}$}          &\multicolumn{3}{c}{$\beta_3\,x_{t3} + \beta_4\,x_{t4}+ \beta_5\,x_{t5}$}       &\multicolumn{3}{c}{$\beta_3\,x_{t3} + \beta_4\,x_{t4}+ \beta_5\,x_{t5}$}      \\                                                         
		models&\multicolumn{3}{c}{}										    &\multicolumn{3}{c}{}                                               &\multicolumn{3}{c}{}                                                           &\multicolumn{3}{c}{$h(\phi_t) = \gamma_1 + \gamma_2\,z_{t2}+$}                 \\                                                       
		&\multicolumn{3}{c}{}  										    &\multicolumn{3}{c}{}                                               &\multicolumn{3}{c}{}                                                           &\multicolumn{3}{c}{$\hskip-0.1truein\gamma_3\,z_{t3} + \gamma_4\,z_{t4} + \gamma_5\,z_{t5}\hskip-0.1truein$}\\ \cmidrule(lr){2-4}   \cmidrule(lr){5-7}    \cmidrule(lr){8-10}    \cmidrule(lr){11-13}     
		$\lambda \rightarrow$&20&50&100&20&50&100&20&50&100&20&50&100\\\hline 
		$P^2$                       &0.759&0.718&0.674&0.545&0.565&0.523&0.638&0.624&0.529&0.885&0.906&0.914\\          
		$P^2_c$                    &0.739&0.695&0.647&0.493&0.515&0.469&0.585&0.569&0.460&0.851&0.878&0.888\\          
		$P^2_{\beta\gamma}$        &0.758&0.716&0.671&0.546&0.567&0.528&0.637&0.624&0.530&0.885&0.906&0.913\\          
		$P^2_{{\beta\gamma}_c}$    &0.738&0.693&0.643&0.494&0.517&0.474&0.584&0.568&0.460&0.851&0.878&0.888\\          
		$R^2_{LR}$                 &0.782&0.743&0.700&0.580&0.611&0.577&0.670&0.653&0.554&0.796&0.816&0.840\\          
		$R^2_{{LR}_c}$             &0.764&0.722&0.675&0.532&0.567&0.529&0.622&0.602&0.488&0.735&0.761&0.792\\\hline    
		\end{tabular}}
	%}
	\end{center}
	\end{table}

\begin{figure}[htp!]
	\centering
%	\vskip-1.0truein
	\makebox{\includegraphics[width=4.0truein,angle=270]
		{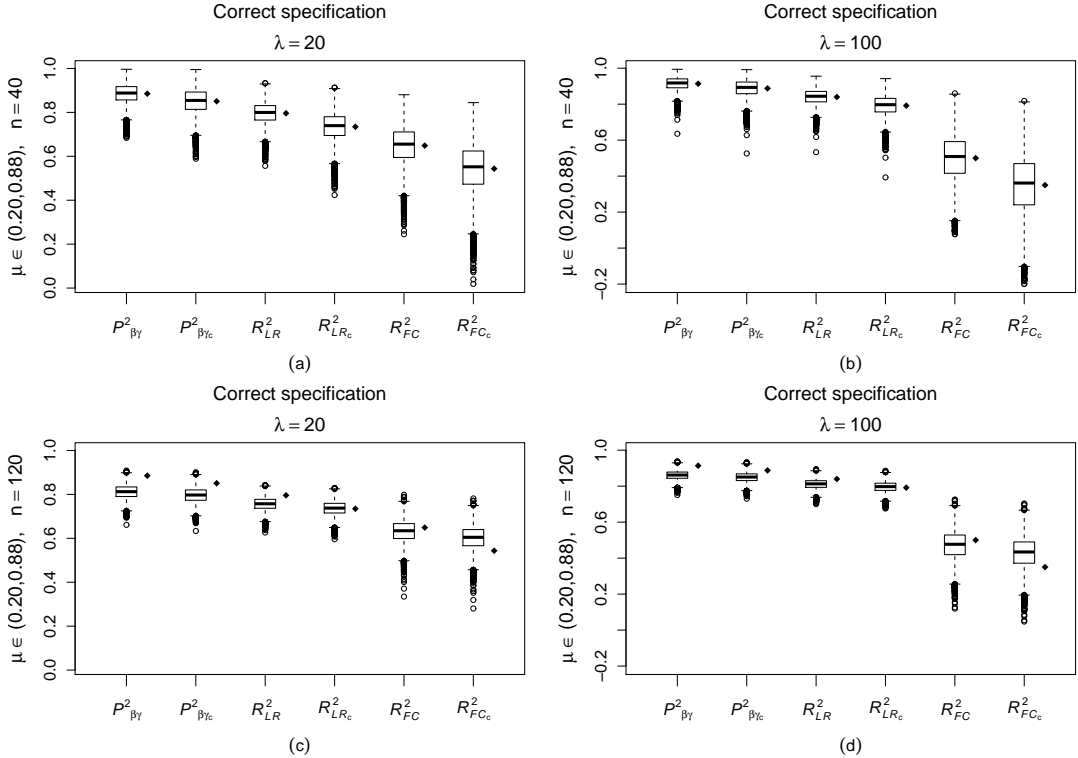}}
%	\vskip-1.2truein
	\caption{\label{fig:4}True model: $g(\mu_t) = \beta_1+{\beta_2}x_{t}$, $h(\phi_t) = \gamma_1+{\gamma_2}z_{t}$.}
\end{figure}

Figure~\ref{fig:4}  summarizes the predictive power performance for each measure. The graphs show that the mean and median performance of $R^2_{FC}$ and $R^2_{FC_{c}}$ are significantly worse than the performance of the other measures. The comparison among the best measures indicate that the median of $P^2$ and $R^2 _{LR}$ performance are significantly better than the measures based on pseudo-$R^2$ and besides reveal some asymmetry of the statistics when the intensity degree of nonconstant dispersion levels increasing. These findings hold in all observation scenarios.
% and at all intensity degree of nonconstant dispersion levels.

%
%The boxplots presented in Figure~\ref{fig:4} confirm the good performance of the statistics $P^2$ and $R^2 _{LR}$
%and besides reveal some asymmetry of the statistics when the sample size increasing. 
%
%
%

\paragraph{Nonlinear models:}
In it follows we shall present Monte Carlo experiments for the class of nonlinear beta regression models. To that end we shall use the  starting values scheme for the estimation by maximum likelihood  proposed by  \cite{ESPINHEIRA2017}. The numerical results were obtained using the following beta regression model as data generating processes: 
$$
\log \left(\frac{\mu_t}{1-\mu_t}\right) = \beta_1+x_{t2}^{\beta_2} + \beta_3{\mbox{log}}(x_{t3} - \beta_4) + \frac{x_{t3}}{\beta_5}, \quad t=1,\ldots,n$$
$x_{t2} \sim U(1,2)$, $x_{t3} \sim U(4.5,34.5)$ and $\phi$ were kept fixed throughout the experiment. The precisions and the sample sizes  are 
$\phi = (20, 50, 150, 400)$, $n = (20, 40, 60, 200, 400).$ Here, the vector of the parameters of the submodel of mean is
$\beta= (1.0,1.9,-2.0,3.4,7.2)^\top$ that produce approximately a range of values for the mean given by $\mu \in(0.36, 0.98).$ To evaluate the performances of statistics on account of nonlinearity negligence we consider the following model specification:
$
\log \left(\frac{\mu_t}{1-\mu_t}\right) = \beta_1+\beta_2 x_{t2} + \beta_3 x_{t3}$. All results are based on 10,000 Monte Carlo replications an for each  replication, we generated the response values as $y_t \sim{\cal B}(\mu_t, \phi_t),$ $t = 1,\ldots, n.$

Table~\ref{T:T3} contains numerical results for the fixed dispersion beta regression model as data generating processes. Here, we compared the performances of the statistics both under incorrect specifications and under correct specification of the nonlinear beta regression model. The results presented in this table reveal that
the $P^2$ and $P^2 _{\beta\gamma}$ statistics outperform the $R^2$ statistics in identifying more emphatically the misspecification. We must emphasize that the response mean is scattered on the standard unit interval. 

Thus, we should not have problems to make prediction and the smaller values of $P^2$
statistics in comparison with the values of the $R^2$ statistics is due to the better performance of the statistics based on residuals in identifying misspecification problems.  For example, fixing the precision on $\phi = 400$, for $n=20$, we have values of $P^2$, $P^2_{\beta\gamma}$, $R^2_{LR}$ and $R^2_{FC}$ equal to $0.576, 0.601, 0.700, 0.637$, respectively. For $n=40$ and  $n=60$ the values of the statistics are $0.568, 0.593, 0.698, 0.634$ and $0.562, 0.588, 0.698, 0.633$, respectively. We can also notice that the values of the  penalized versions of the statistics tend to be greater as the sample size increasing, what it makes sense.

\begin{figure}[!htp]
	\centering
	\vskip-0.35truein
	\makebox{\includegraphics[width=4.3truein,angle=270]{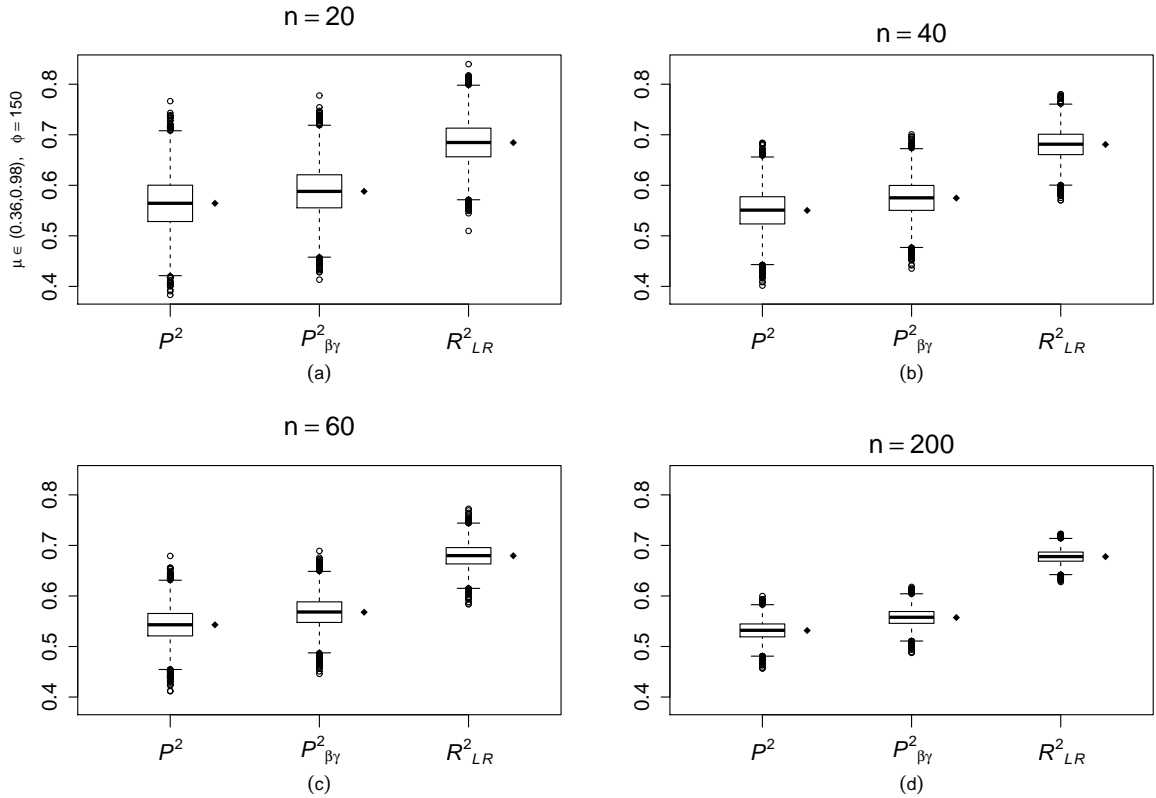}}
	\vskip-0.1truein
	\caption{\label{fig:5} Misspecification: omitted nonlinearity. Estimated model: $g(\mu_t) = \beta_1 + \beta_2\,x_{t2} + \beta_3\,x_{t3}$. True model:
		$g(\mu_t) = \beta_1+x_{t2}^{\beta_2} + \beta_3{\mbox{log}}(x_{t3} - \beta_4) + \frac{x_{t3}}{\beta_5}$, $x_{t2} \sim U(1,2)$, $x_{t3} \sim U(4.5,34.5)$, $\beta= (1.0,1.9,-2.0,3.4,7.2)^\top$, $\,t=1,\ldots,n$, $\phi = 150$, $\mu \in(0.36, 0.98)$.   }
\end{figure}

Figure~\ref{fig:5} summarizes the predictive power measure performance with boxplots over the Monte Carlo replics.  The boxplots clearly show the statistical significance of the performance differences between the measures. The outperformance of the $P^2$ and $P^2_{\beta\gamma}$ statistics in identifying misspecification is more clear when we analyzed the plot.  When the sample size increases, the distributions of the statistics based on residuals tend been concentrated in small values. For the other hand, the distributions of the  $R^2$ statistics tend been concentrated at the same values, considerably greater than the values of the $P^2$ and $P^2 _{\beta\gamma}$ statistics.

\begin{figure}[!htp]
	\centering
	\vskip-0.9truein
	\makebox{\includegraphics[width=5.0truein,angle=270]{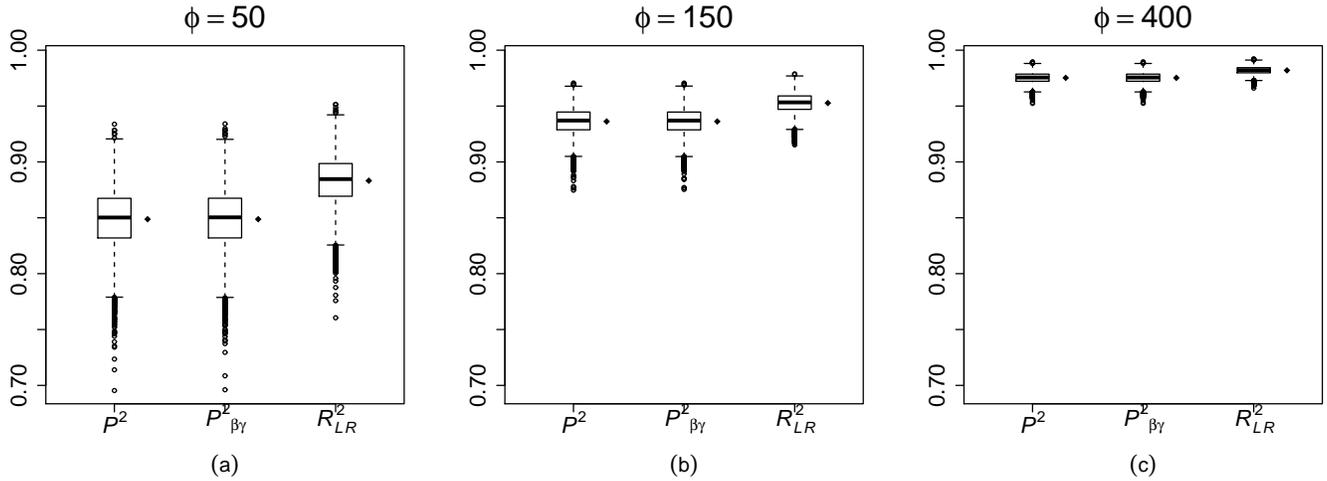}}
	\vskip-1.2truein
	\caption{\label{fig:6} Model correctly specified. True model:
		$g(\mu_t) = \beta_1+x_{t2}^{\beta_2} + \beta_3{\mbox{log}}(x_{t3} - \beta_4) + \frac{x_{t3}}{\beta_5}$,  $\,t=1,\ldots,n$,   $x_{t2} \sim U(1,2)$, $x_{t3} \sim U(4.5,34.5)$, $\beta= (1.0,1.9,-2.0,3.4,7.2)^\top.$ Range of values for mean 
%		$n = 60$
		 $\mu \in(0.36, 0.98)$. }
	\end{figure}

Figure~\ref{fig:6} summarizes the empirical distribution behavior of predictive power measure when $n=60$. The graphs show that the median performance of $R^2_{LR}$ is significantly worse than the performance of the $P^2$ and $P^2_{\beta\gamma}$ measures. However, under  true specification  the statistics perform equally well and as the precision of the model increases the values of the statistics tend being concentrated close to one. Also, we notice that the performance comparison among different levels of $\phi$ shows a systematic increase of the power performance.

	\begin{table}[htp!]
	\begin{center}
		\caption{\label{T:T3} Values of the statistics. True model:
			$g(\mu_t) = \beta_1+x_{t2}^{\beta_2} + \beta_3{\mbox{log}}(x_{t3} - \beta_4) + \frac{x_{t3}}{\beta_5}$, $x_{t2} \sim U(1,2)$, $x_{t3} \sim U(4.5,34.5)$, $\beta= (1.0,1.9,-2.0,3.4,7.2)^\top$, $\mu \in(0.36, 0.98)$, $\,t=1,\ldots,n$, $\phi$ fixed.
			Misspecification: : $g(\mu_t) = \beta_1 + \beta_2\,x_{t2} + \beta_3\,x_{t3}$ (omitted nonlinearity).
			}
			\medskip
		\renewcommand{\tabcolsep}{0.35pc} % enlarge column spacing
		\renewcommand{\arraystretch}{2.1}
		%\smallskip
		%\centering
		%\fbox{%
		%\begin{tabular}{@{}|c|c|c|c|c|c|c|c|c|c|c|c|c|} \hline
		{\tiny
			\begin{tabular}{*{16}{c}} \hline
%				Model:&\multicolumn{15}{c}{$\mu \in(0.36, 0.98)$}\\\hline
%				True Model        &\multicolumn{15}{c}{$g(\mu_t) = \beta_1+x_{t2}^{\beta_2} + \beta_3{\mbox{log}}(x_{t3} - \beta_4) + \frac{x_{t3}}{\beta_5}$}\\\hline
				Estimated Model   &\multicolumn{12}{c}{With misspecification: $g(\mu_t) = \beta_1 + \beta_2\,x_{t2} + \beta_3\,x_{t3}$}&\multicolumn{3}{c}{Correctly}\\ \cmidrule(lr){2-13} \cmidrule(lr){14-16}
				n&\multicolumn{4}{c}{20}&\multicolumn{4}{c}{40}&\multicolumn{4}{c}{60}&\multicolumn{3}{c}{60}\\ \cmidrule(lr){2-5} \cmidrule(lr){6-9} \cmidrule(lr){10-13} \cmidrule(lr){14-16}
				$\phi \rightarrow$&20&50&150&400&20&50&150&400&20&50&150&400&50&150&400\\\hline 
	   $P^2$                      &0.485& 0.535& 0.564& 0.576& 0.438&	0.508&0.550&0.568&0.420&0.496&0.543&0.562&0.849&0.936&0.975\\            
	   $P^2_c$                    &0.388& 0.448& 0.483& 0.497& 0.391&	0.467&0.513&0.532&0.388&0.469&0.518&0.539&0.835&0.930&0.973\\            
	   $P^2_{\beta\gamma}$        &0.502& 0.556& 0.588& 0.601& 0.456&	0.531&0.575&0.593&0.439&0.520&0.568&0.588&0.849&0.936&0.975\\            
	   $P^2_{{\beta\gamma}_c}$    &0.409& 0.473& 0.511& 0.526& 0.411&	0.492&0.539&0.559&0.409&0.494&0.545&0.566&0.835&0.930&0.973\\            
	   $R^2_{LR}$                 &0.578& 0.647& 0.684& 0.700& 0.563&	0.639&0.681&0.698&0.557&0.636&0.680&0.698&0.883&0.953&0.982\\            
	   $R^2_{{LR}_c}$             &0.499& 0.581& 0.625& 0.643& 0.526&	0.608&0.654&0.673&0.533&0.616&0.662&0.681&0.863&0.945&0.979\\            
	   $R^2_{FC}$                 &0.486& 0.574& 0.619& 0.637& 0.448&	0.556&0.612&0.634&0.437&0.550&0.609&0.633&0.879&0.951&0.981\\            
	   $R^2_{{RC}_c}$             &0.389& 0.494& 0.548& 0.569& 0.402&	0.519&0.580&0.604&0.407&0.526&0.588&0.613&0.867&0.946&0.979\\\hline      
	   				\end{tabular}}
			%}
			\end{center}
		\end{table}

\paragraph{Nonlinearity on dispersion model:}
The last simulations consider two nonlinear submodels both to mean and dispersion, namely:
$$\log \left(\frac{\mu_t}{1-\mu_t}\right) = \beta_1+x_{t}^{\beta_2}\quad \text{and}\quad \log \left({\phi_t}\right) = \gamma_1+z_{t}^{\gamma_2}.$$ 		

\eject		
		
%\eject

\begin{figure}[!htp]
	\centering
%	\vskip-0.5truein
	\makebox{\includegraphics[width=4.0truein,angle=270]
		{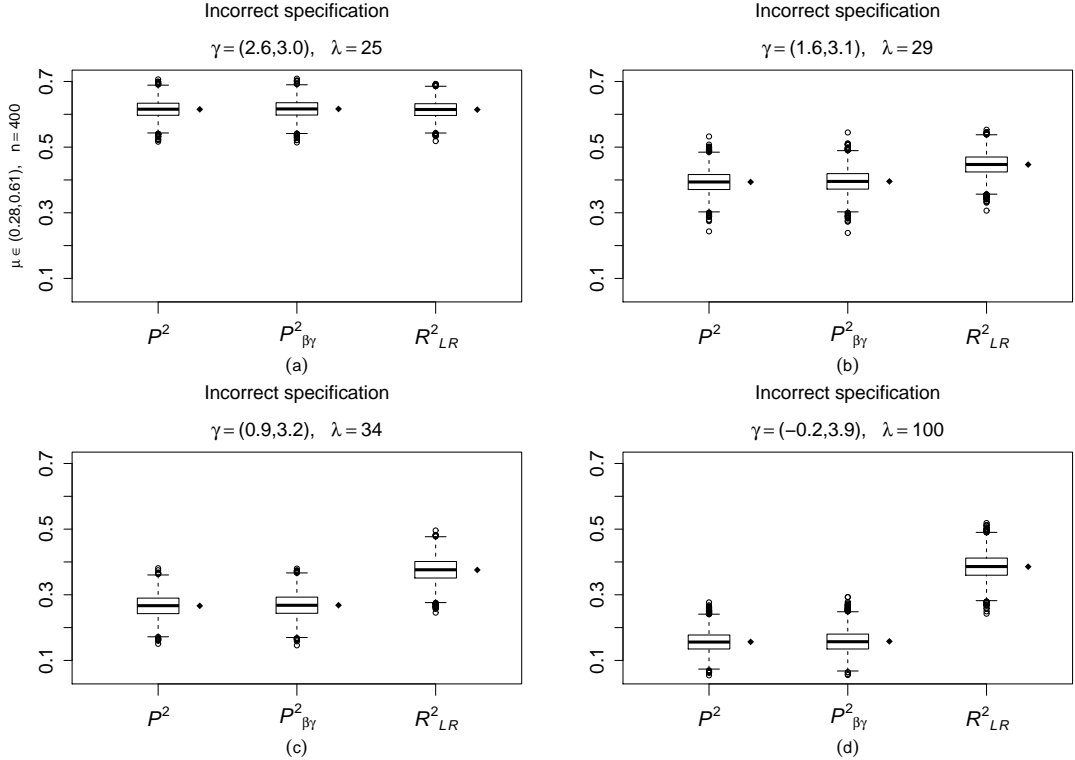}}
%	\vskip-1.1truein
	\caption{\label{fig:7} Misspecificated model: $g(\mu_t) = \beta_1+{\beta_2}x_{t}$, $h(\phi_t) = \gamma_1+{\gamma_2}z_{t}$. True model:
		$g(\mu_t) = \beta_1+x_{t}^{\beta_2}$, $h(\phi_t) = \gamma_1+z_{t}^{\gamma_2}$, $x_{t} \sim U(0.3,1.3)$, $z_{t} \sim U(0.5,1.5)$, $\beta= (-1.1,1.7)^\top$,  $\,t=1,\ldots,n$, $\mu \in(0.28, 0.61)$, $n=400$. }
\end{figure}

%\paragraph{Nonlinearity on dispersion model:}
%The last simulations consider two nonlinear submodels both to mean and dispersion, namely:
%$$\log \left(\frac{\mu_t}{1-\mu_t}\right) = \beta_1+x_{t}^{\beta_2}\quad \text{and}\quad \log \left({\phi_t}\right) = \gamma_1+z_{t}^{\gamma_2}.$$ 
We fixed: $n=400$,
$\beta= (-1.1,1.7)^\top$, $x_{t} \sim U(0.3,1.3)$; ($\mu \in(0.28, 0.61)$), $z_{t} \sim U(0.5,1.5)$ and  we
varying $\gamma$ such that $\gamma= (2.6,3.0)^\top$; $\lambda \approx 25$, $\gamma= (1.6,3.1)^\top$; $\lambda \approx 29$, 
$\gamma= (0.9,3.2)^\top$; $\lambda \approx 35$ and $\gamma= (-0.3,3.9)^\top$; $\lambda \approx 100$,  
$\,t=1,\ldots,n$.  In Figure~\ref{fig:7} we present the boxplots of the $P^2$, $P^2_{\beta\gamma}$ and $R^2_{LR}$ under negligence of nonlinearity, that is the estimated model is 
$\log \left(\frac{\mu_t}{1-\mu_t}\right) = \beta_1+{\beta_2} x_{t}\quad \text{and}\quad \log \left({\phi_t}\right) = \gamma_1+{\gamma_2}z_{t}$. Based on this figure we notice that once again the statistics based on residuals outperform the $R^2$ statistics since that the values of the $P^2$ and $P^2_{\beta\gamma}$ are considerably smaller than the values of the $R^2_{LR}$ statistic, in especial when the nonconstant dispersion is more several (when $\lambda$ increases). 

\begin{figure}[htp!]
	\centering
%	\vskip-0.4truein
	\makebox{\includegraphics[width=3.0truein,angle=270]{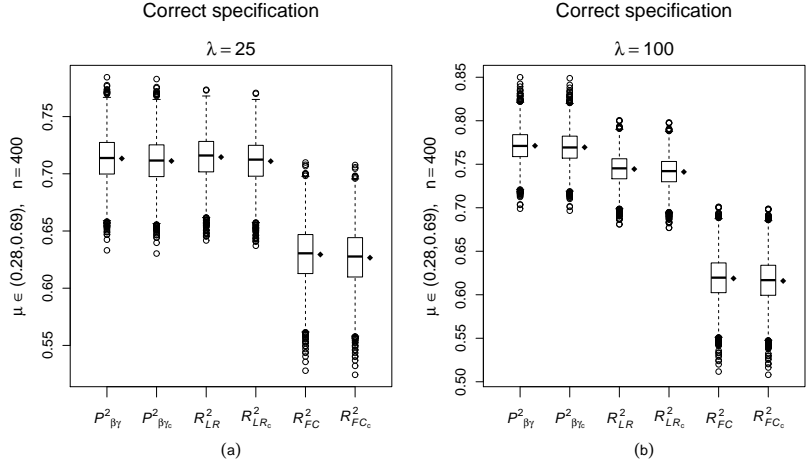}}
	\vskip-0.4truein
	\caption{\label{fig:8} True model:
		$g(\mu_t) = \beta_1+x_{t}^{\beta_2}$, $h(\phi_t) = \gamma_1+z_{t}^{\gamma_2}$, $x_{t} \sim U(0.3,1.3)$, $z_{t} =x_{t}$, $\beta= (-1.1,1.7)^\top$, $\gamma= (2.3,5.3)^\top$, $n=400$. }
\end{figure}

Figure~\ref{fig:8} summarizes the predictive power measure performance with boxplots over the Monte Carlo replics. Here, we evaluated the empirical distribution of the $R^2_{FC}$ statistics under nonconstant dispersion for two  models estimated correctly. The plots 
reveals evidences that the pseudo-$R^2$  measures are not a good statistics for  model selection when the dispersion varying along the observations. It is clear that $P^2$ and $P^2_{\beta\gamma}$ measures become more powerful as the $\lambda$ increases.

\section{Applications}
In what follows we shall present an application based on real data. 
\paragraph{Application I:} 
The application relates to the distribution of natural gas for home usage
(e.g., in water heaters, ovens  and stoves) in S\~ao Paulo, Brazil. 
Such a distribution is based on two factors: the simultaneity factor ($F$) and   the total nominal power  of appliances that use natural gas, computed power $Q_{max}$.
Using these factors one obtains an indicator of gas release in a given tubulation section, namely:
$Q_p = F \times Q_{max}$.  The simultaneity factor  assumes values in $(0,1)$, and can be interpreted as the probability of simultaneous appliances usage.
Thus, based on $F$ the company that supplies the gas  decides how much gas
to supply to a given residential unit.

%
%When the lager $\lambda$ the  $P^2$ and $P^2_{\beta\gamma}$ measures performs better compared to the other methods 
%
%The Figure~\ref{fig:8} presents the results for models estimated correctly. Here, we also evaluated the empirical distribution of the $R^2_{FC}$ statistics under nonconstant dispersion. Once more the Figure~\ref{fig:8} reveals evidences that the $R^2_{FC}$ is not a good statistic for  model selection when the dispersion varying along the observations.
%

\eject

The data were analyzed by \cite{Zerbinatti2008}, obtained from  the Instituto de Pesquisas Tecnol\'ogicas (IPT) and the Companhia de G\'as de S\~ao Paulo (COMG\'AS).  The response variable (y) are the simultaneity factors of 42 valid measurements of sampled households, and the covariate is the computed power ($x_1$).  The simultaneity factors ranged
from 0.02 to 0.46, being the  median   equals 0.07. \cite{Zerbinatti2008} modeled such data and concluded
that the best performing model was  the beta regression model based on logit link function for the mean and logarithmic form (log) of  computed power used as covariate.
However, the author shows that the beta regression model can underpredict the response. Thus, \cite{Espinheira2014} argue that it is important to have at disposal prediction intervals that can be used with beta regressions. To that end, the authors built and evaluated bootstrap-based prediction intervals for the response for the class of beta regression models.  They applied the approach to the data on simultaneity factor. However, a important step in this case was the selection of the model with the best predictive power. 
To reach this aim the authors used a  simplified version of PRESS statistic given by 
$PRESS ={ {\sum_{t=1}^{42}(y_t - \widehat y_{(t)})^2} / { 42 }}$ which selected the same model of
\cite{Zerbinatti2008}.
Here we aim at selecting the better  predictive model to the data on simultaneity factor using the $P^2$ and $P^2_{\beta\gamma}$ statistics.
We also consider the $R^2_{LR}$ and $R^2_{RC}$ as  measures of goodness-of-fit model.
The response is the simultaneity factor and the covariate $X_{2}$ is the log  of
computed power. 
We adjusted four  beta regression models, considering constant and nonconstant dispersion and  logit  or log-log  link function for $\mu$. For the varying dispersion model we used the log link function for $\phi$. 
The  values of the statistics are presented in Table~\ref{T:T4}.
Here, we should emphasize that the model predictive power is better when the measures $P^2$ and $P^2_{\beta\gamma}$ are close to one.

\begin{table}[!htp]
\begin{center}
	\caption{\label{T:T4} Values of the statistics from the candidate models. Data on simultaneity factor} 
	%\centering
	%\fbox{%
	{\scriptsize
			\renewcommand{\tabcolsep}{0.35pc} % enlarge column spacing
			\renewcommand{\arraystretch}{1.5}
		\begin{tabular}{*{5}{c}}\hline
			&\multicolumn{4}{c}{\text{Candidate models}}\\\hline
			Mean&{$\log( {\mu_t}/{(1-\mu_t)}) = $}&{$-\log(-\log {(\mu_t)}) = $}&{$\log( {\mu_t}/{(1-\mu_t)}) = $}&{$-\log(-\log {(\mu_t)}) = $}\\ submodel&{$\beta_1 + \beta_2\,x_{t1}$}&{$\beta_1 + \beta_2\,x_{t1}$}&{$\beta_1 + \beta_2\,x_{t1}$}&{$\beta_1 + \beta_2\,x_{t1}$}\\\hline
			Dispersion& \multirow{ 2}{*}{--} & \multirow{ 2}{*}{--} &{$\log(\phi_t) = $}&{$\log(\phi_t) = $}\\ 
			submodel&&&{$\gamma_1 + \gamma_2\,x_{t1}$}&{$\gamma_1 + \gamma_2\,x_{t1}$}\\\hline
			$P^2$                      &0.66&	   0.42&  0.70&  0.88\\       
			$P^2_c$                    &0.64&	   0.39&  0.68&  0.87\\       
			$P^2_{\beta\gamma}$        &0.65&	   0.42&  0.70&  0.88\\       
			$P^2_{{\beta\gamma}_c}$    &0.64&	   0.39&  0.68&  0.87\\       
			$R^2_{LR}$                 &0.72&	   0.70&  0.74&  0.74\\       
			$R^2_{{LR}_c}$             &0.69&	   0.65&  0.70&  0.70\\
			$R^2_{FC}$                 &0.69&	   0.72&  0.70&  0.72\\       
			$R^2_{{FC}_c}$             &0.67&	   0.70&  0.67&  0.70\\
			\hline 
		\end{tabular}}
		%}
		\end{center}
	\end{table}
	
The Table~\ref{T:T4} displays two important informations.  First, we notice that by the $R^2$ measures  the models equally fits well.  Second, the $P^2$ and $P^2_{\beta\gamma}$ measures lead to the same conclusions,  selecting  the beta regression model with link log-log for the mean submodel and link log for the dispersion submodel,   as the best model to make prediction to the data on   simultaneity factor.
The maximum 
likelihood parameter estimates are   $\widehat{\beta}_1 = -0.63$, $\widehat{\beta}_2 = -0.31$, $\widehat{\gamma}_1 = 3.81$ and $\widehat{\gamma}_2 = 0.77$.
Furthermore, the estimative of  intensity of nonconstant dispersion  is $\widehat\lambda = 21.16$ (see (16)), such that $\widehat\phi_{\max} = 242.39$ and  $\widehat \phi_{\min} = 11.45$. Selected among the candidates  the best model in a predictive perspective, we still can use  the PRESS statistic to identifying which observations are more difficult to predict.  In this sense, we plot the individual components of PRESS$_{\beta\gamma}$ versus the observations index and we added a horizontal line at  $3\sum_{t=1}^n{\text{PRESS}}_{{\beta\gamma}_t}/n$ and singled out points that considerably
exceeded this threshold. 

\begin{figure}[!htp]
\centering
\makebox{\includegraphics[width=3.0truein,angle=270]{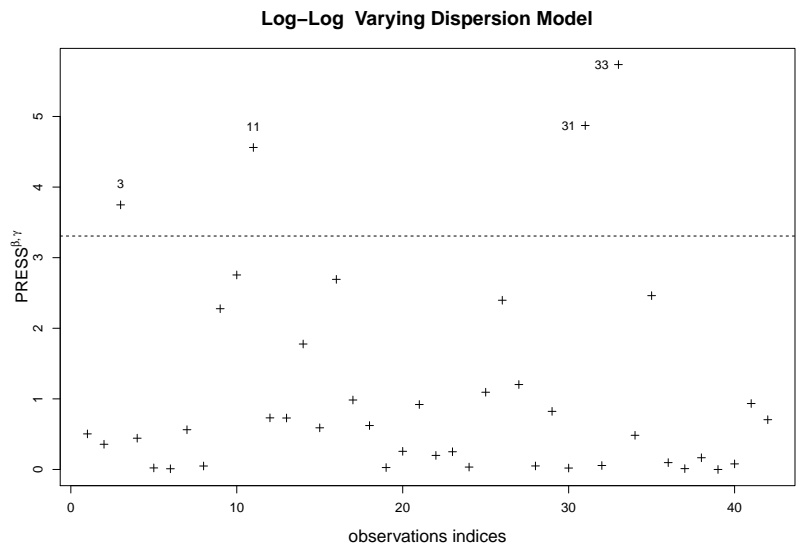}}
\caption{\label{fig:9} PRESS plot. Data on simultaneity factor}
\end{figure}

Figure~\ref{fig:9} shows that the cases 
3, 11, 33 and 33 arise as the observations with more predictive difficulty and are worthy of further investigation.

\paragraph{Application II:} The second application consider a nonlinear beta regression used to modeling the proportion of killed grasshopper ($y$) at an assays on a grasshopper Melanopus sanguinipes with the insecticide carbofuran and the synergist piperonyl butoxide. This model was proposed by \cite{ESPINHEIRA2017} after careful building the scheme of starting values to iterative process of maximum likelihood estimation and after a meticulous residual analysis. Our aim here is applies both predictive power statistics and goodness-of-fit statistics to confirm or not  the choice of the model made by residual analysis. The covariates are the dose of the insecticide ($x_1$) and the dose of the synergist ($x_2$). The data can be found in \citet[p.\ 385]{MCCULLAGH1989}. Additionally, $y\in[0.04,0.84]$, $\mu_{y} = 0.4501$ and the median of the response is equal to $0.4967$.

The model selected with its estimates and respective p-values is present in
\cite{ESPINHEIRA2017} and is given by
$\log( {\mu_t}/{1-\mu_t})= \beta_1 + \beta_2 \log{(x_{t1}- \beta_3)} + \beta_4 \frac{x_{t2}}{x_{t2}+\beta_5}$ and 
	$\sqrt{\phi_t}= \gamma_1 + \gamma_2 x_{t1} + \gamma_3 x_{t2}$, $t=1,\ldots,15$.
Now, by using  residual analysis as diagnostic tools several linear beta regression models are compared with  nonlinear model. After competition the nonlinear model:  	
%	
%	
%Here, we investigated, using residual analysis, several linear beta regression models and the final candidate to compete with the nonlinear model was:
$\log ({\mu_t}/{1-\mu_t})= \beta_1 + \beta_2 \log{(x_{t1} + 1.0)} + \beta_3 x_{t2}$
	$\sqrt{\phi_t}=\gamma_1+\gamma_2 x_{1t}$
$t=1, \ldots, 15$ was selected. The estimatives of parameters are
$\widehat\beta_1 = -4.25;\widehat\beta_2=1.79;\,\widehat\beta_3=0.04;\,
\widehat\gamma_1=1.19$ and $\widehat\gamma_2=0.19$. The values of the $P^2$ and $R^2$ measures for the two candidate models are present Table \ref{T:T5}. Based on this table we can note that the nonlinear model outperforms the linear model under all statistics, that is, the nonlinear model has both better predictive power and better goodness-of-fit. 

\begin{figure}[!htp]
	\centering
	\makebox{\includegraphics[width=3.0truein,angle=270]{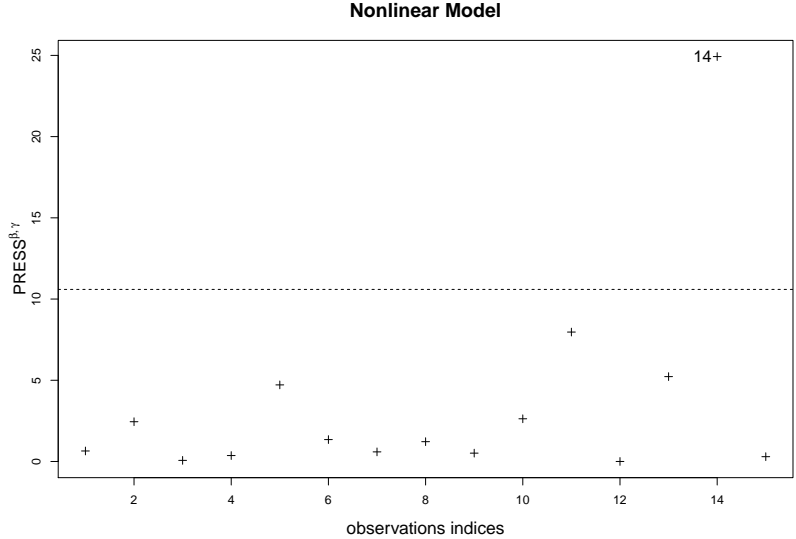}}
	\caption{\label{fig:10} PRESS plots. Data on insecticide.}
\end{figure}

With aim in identifying observations for which to make prediction can be a hard task we plot values of PRESS statistic versus indices of the observations. In  Figure~\ref{fig:10} it is noteworthy how the case 14 is strongly singled out. In fact, this case was also singled out in plots of residual analysis made by \cite{ESPINHEIRA2017}. However, the observation 14 is not an influential case, in sense of to affect inferential results. Besides, the choose model was capable to estimated well this case.

Thus, we confirm by the model selection measures that the beta nonlinear model proposed by \cite{ESPINHEIRA2017} is a suitable alternative to modeling of the data of insecticide carbofuran and the synergist piperonyl butoxide \citet[p.\ 385]{MCCULLAGH1989}.
\begin{table}[!htp]
\begin{center}
	\caption{\label{T:T5} Values of the statistics from the candidate models. Data on insecticide.} 
		\vskip0.14truein
	\centering
	%\fbox{%
	{\scriptsize
			\renewcommand{\tabcolsep}{0.35pc} % enlarge column spacing
				\renewcommand{\arraystretch}{1.5}
		\begin{tabular}{*{3}{c}}\hline
			&\multicolumn{2}{c}{\text{Candidate models}}\\\hline
			&{\text{Linear Models}}&{\text{Nolinear models}}\\\hline
			Mean&{$\log( {\mu_t}/{(1-\mu_t)}) = $}&{$\log( {\mu_t}/{(1-\mu_t)}) = $}
			\\ submodel&{$\beta_1 + \beta_2 \log{(x_{t1} + 1.0)} + \beta_3 x_{t2}$}&{$\beta_1 + \beta_2 \log{(x_{t1}- \beta_3)} + \beta_4 \frac{x_{t2}}{x_{t2}+\beta_5}$}\\\hline
			Dispersion&{$\sqrt{\phi_t}= $}&{$\sqrt{\phi_t} = $}\\ 
			submodel&{$\gamma_1 + \gamma_2x_{t1}$}&{$\gamma_1+\gamma_2 x_1 + \gamma_3 x_2+ \gamma_4 (x_1 x_2)$}\\\hline
			$P^2$                      &	   0.89&    0.99\\       
			$P^2_c$                    &	   0.85&    0.99\\       
			$P^2_{\beta\gamma}$        &	   0.89&    0.99\\       
			$P^2_{{\beta\gamma}_c}$    &	   0.86&    0.99\\       
			$R^2_{LR}$                 &	   0.83&    0.99\\       
			$R^2_{{LR}_c}$             &	   0.70&    0.99\\
			$R^2_{FC}$                 &	   0.79&    0.97\\       
			$R^2_{{FC}_c}$             &	   0.71&    0.94\\
			\hline 
		\end{tabular}}
		%}
			\end{center}
	\end{table}

\paragraph{Application III:}  In the latter application we will use the dataset about available chlorine fraction    after weeks of
manufacturing  from an investigation performed at Proctor \& Gamble. A certain product must have a fraction of
available chlorine equal to 0.50 at the time of manufacturing. It is known that chlorine fraction
of the product decays with time. Eight weeks after the production, before the product is
consumed, in theory there is a decline to a level 0.49.

The theory related to the problem indicates that the available chlorine fraction ($y$)
decays according to a nonlinear function of the number of weeks ($x$) after fabrication of the product and
unknown parameters (Draper and Smith, 1981 p. 276), given by
\begin{equation}\label{eq9}
\eta_t = \beta_1 +(0.49 - \beta_1){\rm exp}\{\beta_2(x_{t}-8)\}.
\end{equation}
The level 0.49 depends on several uncontrolled factors, as for example warehousing environments or handling facilities. Thus, the predictions based on theoretical model can be not reliable.

Cartons of the product were analyzed over a period aiming answer some questions 
like as: {\it``When should warehouse material be scrapped?"}
or {\it ``When should store stocks be replaced?"} According to knowledgeable chemists an equilibrium asymptotic level of available chlorine should be expected somewhere close to  0.30.

From predictor based on \eqref{eq9} we can note that when $x=8$ the nonlinear model provides a true level for the available chlorine fraction (no error), wherein $\eta=0.49$. We consider a new logit nonlinear beta regression model. We replaced the deterministic value 0.49 by an additional parameter at the predictor. Thus, the new nonlinear  predictor for mean submodel is given by
$\eta_t = \beta_1 +(\beta_3 - \beta_1){\rm exp}\{-\beta_2(x_{t}-8)\}$, $t=1,\ldots,42$.
Here the available chlorine fraction ranged
from 0.38 to 0.49, being the mean and median are approximately equal 0.42. We investigated some competitive models. Table~\ref{T:T6} the results of 
final candidates models and its statistics. The findings reveals that the model $\log( {\mu_t}/{(1-\mu_t)}) = \beta_1+(\beta_3 - \beta_1){\rm exp}\{\beta_2(x_{t1}-8)$ and $
\log(\phi_t) = \gamma_1 +\gamma_2\mbox{log}x_{t1} + {\rm exp}\{\gamma_3(x_{t1}-8)$ is the best performer in sense that it displays the higher statistics values. To estimate  this model was necessary  to build a starting values procedure for log-likelihood maximization as  proposed by \cite{ESPINHEIRA2017}. Since that we have more parameters than covariates firstly we  used the theoretical information about the asymptotic level and found a initial guess to $\beta_1$ equal to 0.30. Thus, based on equation we  took some values to $y$ and $x_1$  and found  a initial  guess to $\beta_2$, $\beta_2^{(0)} = 0.02$. Then we carried out the scheme of starting values to be
used in nonlinear beta regression maximum likelihood estimation \citep{ESPINHEIRA2017}. 
The parameters estimatives are $\widehat\beta_1 =-0.45963 $  $\widehat\beta_2 =-0.04166$  $\widehat\beta_3 =0.09479$ $\widehat\gamma_1 =13.18335$     $\widehat\gamma_2 =-0.05413$  -$\widehat\gamma_3 =2.63158$. It is importance emphasize that the  $\beta_2$ estimative conduce to a level of 
chlorine fraction equal to $0.4896\approx0.49$ that for this dataset  confirm the theory that there is a decline to a level 0.49.

\begin{table}[!htp]
\begin{center}
	\caption{\label{T:T6} Values of the statistics from the candidate models. Data on chlorine fraction} 
	\renewcommand\arraystretch{1.3}% enlarge row spacing
	\renewcommand{\tabcolsep}{0.19pc} % enlarge column 
	%\centering
	%\fbox{%
	{\scriptsize
			\renewcommand{\tabcolsep}{0.2pc} % enlarge column spacing
					\renewcommand{\arraystretch}{1.5}
		\begin{tabular}{*{5}{c}}\hline
			&\multicolumn{4}{c}{\text{Candidate models}}\\\hline
		\multirow{ 2}{*}{$\eta_{1t}$} 	&{$\log( {\mu_t}/{(1-\mu_t)}) = \beta_1+ $}&{$-\log(-\log {(\mu_t)}) = \beta_1+$}&{$\log( {\mu_t}/{(1-\mu_t)}) = \beta_1+ $}&{$-\log(-\log {(\mu_t)}) = \beta_1+$}\\
			&{$(\beta_3 - \beta_1){\rm exp}\{\beta_2(x_{t1}-8)\}$}&{$(\beta_3 - \beta_1){\rm exp}\{\beta_2(x_{t1}-8)\}$}&{$(\beta_3 - \beta_1){\rm exp}\{\beta_2(x_{t1}-8)\}$}&{$(\beta_3 - \beta_1){\rm exp}\{\beta_2(x_{t1}-8)\}$}\\\hline
			\multirow{ 2}{*}{$\eta_{2t}$}& \multirow{ 2}{*}{--}                                          &     \multirow{ 2}{*}{--}                         &{$\log(\phi_t) = \gamma_1 +\gamma_2\mbox{log}x_{t1} $}&{$\log(\phi_t) = \gamma_1 + \gamma_2\mbox{log}x_{t1}$}\\ 
			&&&{$ + {\rm exp}\{\gamma_3(x_{t1}-8)\}$}&{$ + {\rm exp}\{\gamma_3(x_{t1}-8)\}$}\\\hline
			$P^2$                      &0.88&	   0.87&  0.96&  0.95\\       
			$P^2_c$                    &0.87&	   0.86&  0.96&  0.95\\       
			$P^2_{\beta\gamma}$        &0.88&	   0.87&  0.96&  0.95\\       
			$P^2_{{\beta\gamma}_c}$    &0.87&	   0.86&  0.96&  0.95\\       
			$R^2_{LR}$                 &0.87&	   0.87&  0.88&  0.88\\       
			$R^2_{{LR}_c}$             &0.85&	   0.85&  0.86&  0.86\\\hline 
		\end{tabular}}
		%}
		\end{center}
	\end{table}

\section{Conclusion}
In this paper we develop the   $P^2$ and  $P^2 _ {\beta\gamma}$ measures  based on two versions of PRESS 
statistics for the class of beta regression models.   The $P^2$ coefficient consider 
the PRESS statistic  
based on ordinary residual obtained from the Fisher's scoring iterative algorithm for estimating $\beta$
whereas $P^2 _ {\beta\gamma}$ is based on a new residual which is a combination of ordinaries residuals from the Fisher's scoring iterative algorithm for estimating $\beta$ and $\gamma$.
We have  presented the results of  Monte Carlo simulations carried out to evaluate the performance of predictive coefficients. Additionally, to access the goodness-of-fit model we used the $R^2_{LR}$ and $R^2_{FC}$ measures. We consider different scenarios including misspecification of omitted covariates and  negligence of varying dispersion,  simultaneous increase in the number of covariates in the two submodels (mean and dispersion) and and  negligence of nonlinearity. 

In general form, the  coefficients $P^2$ and  $P^2 _{\beta\gamma}$ perform similar and both enable to identify when the  model are not reliable to predict or when is more difficult to make prediction. In particular it is noteworthy that when the response values are close to one or close to zero the  power predictive of the model is substantially affected even under correct specification. In these
situations, the  $R^2$ statistics also revel that the model does not fit well.

Other important conclusion is about the bad performance of the $R^2{FC}$ for beta regression models with varying dispersion, in sense that even when the model is well specified the values of this statistic tend to be too smaller than the values of the others statistics.

Finally, three empirical applications were performed and yield to a relevant information. In cases that the $R^2$ statistics evaluate the candidate models quite equally the predictive measures were decisive to choose the best between the candidate models. But as suggestion, to selected a model even in a predictive sense way it is also important to use goodness-of-fit measures and our recomendation  for the class of nonlinear models is to use the version of $R^2 _{LR}$ considered by \cite{BAYER2017} as the more appropriated  model select criteria to linear beta regression models with varying dispersion.

\section*{Acknowledgement}
	This work was supported in part by Conselho Nacional de Desenvolvimento Cient{\'i}fico
	e Tecnol{\'o}gico (CNPq) and Funda\c{c}{\~a}o de Amparo {\`a} Ci{\^e}ncia e Tecnologia de Pernambuco (FACEPE).
%\end{acknowledgement}
%\vspace*{1pc}

% \section*{Acknowledgments}
% This work was supported in part by Conselho Nacional de Desenvolvimento Científico
% e Tecnológico (CNPq) and Fundação de Amparo à Ciência e Tecnologia de Pernambuco (FACEPE).

\section*{Conflict of Interest}

The authors have declared no conflict of interest. 

% \section*{Appendix {\it(please insert here, if applicable)}}
\appendix
\section*{Appendix}
\label{apendice}
{\footnotesize
%\noindent 
\paragraph{Fisher's scoring iterative algorithm:} In what follows we shall present the score function and Fisher's information for $\beta$ and $\gamma$ in nonlinear beta regression models \citep{SIMAS2010}. The log-likelihood function for model \eqref{eq1} is given by $
	\ell (\beta, \gamma)=\sum_{t=1}^{n} \ell_t(\mu_t, \phi_t),
$ and $
	\ell_t(\mu_t, \phi_t)=\log\Gamma(\phi_t)-\log\Gamma(\mu_t\phi_t)-\log \Gamma((1-\mu_t)\phi_t)+(\mu_t\phi_t-1)\log y_t
	+\{(1-\mu_t)\phi_t-1\}\log(1-y_t).  
$
The score function for $\beta$ is 
\begin{equation}
\label{A1}
U_{\beta}(\beta, \gamma) =  J^\top_1\Phi T (y^*-\mu^{*}),
\end{equation}
where $J_1=\partial \eta_{1}/\partial \beta$ (an $n \times k$ matrix), $\Phi = {\rm diag}\{\phi_1,\ldots,\phi_n\}$, the $t$th elements of $y^*$ and $\mu^*$ being given in \eqref{eq5}. Also,  
$	T = {\rm diag}\{ 1/g'(\mu_1), \ldots, 1/g'(\mu_n)\}.
$. The score function for $\gamma$ can be written as 
$U_{\gamma}(\beta, \gamma) =  J^\top_2Ha,$
where $J_2=\partial \eta_2/\partial \gamma$ (an $n\times q$ matrix), $a_t$ is given in~\eqref{eq7} and 
$	H = {\rm diag}\{ 1/h'(\phi_1), \ldots, 1/h'(\phi_n)\}. 
$
The components of Fisher's information matrix are
\begin{equation}
\label{A2}
K_{\beta\beta} = J^\top_1\Phi WJ^\top_1,\quad 
K_{\beta\gamma}
= K_{\gamma\beta}^{\!\top} = J^\top_1CTHJ^\top_2\quad {\text and}\quad K_{\gamma\gamma} = J^\top_2 D J^\top_2.
\end{equation}
Here, $W = {\rm diag}\{ w_1, \ldots, w_n\}$, where
\begin{equation}
	\label{A3}
	w_t = \phi_t v_t [1 / \{g'(\mu_t)\}^2] \quad \text{and}\quad v_t = \left\{ \psi'(\mu_t\phi_t) + \psi'((1-\mu_t)\phi_t)\right\}.
\end{equation}
Also, $C = {\rm diag}\{ c_1, \ldots, c_n\}$;
$c_t = \phi_t \left\{ \psi'(\mu_t\phi_t)\mu_t - \psi'((1-\mu_t)\phi_t)
(1-\mu_t)\right\}$,
$D={\rm diag}\{ d_1, \ldots, d_n\}$; 
$d_t = \xi_t /(h'(\mu_t))^2$ and $\xi_t = \left\{\psi'(\mu_t\phi_t)\mu_t^2 + \psi'((1-\mu_t)\phi_t)(1-\mu_t)^2
	- \psi'(\phi_t)\right\}.$
To propose PRESS statistics for a beta regression we shall based on Fisher iterative maximum likelihood scheme and weighted least square regressions. 
Fisher's scoring iterative scheme used for estimating $\beta$, both to linear and nonlinear regression model,  can be written as 
\begin{equation}
	\label{A4}
	\beta^{(m+1)} =\beta^{(m)} +  {(K_{\beta\beta}^{(m)})^{-1}}U_{\beta}^{(m)}(\beta). \quad \end{equation}
Where $ m = 0,1,2,\ldots$ are the iterations
which are carried out until convergence.
The convergence happens when the difference
$|\beta^{(m+1)}-\beta^{(m)}|$ 
is less than a small, previously specified constant.

From \eqref{A1}, \eqref{A2} and \eqref{A4} it follows that the $m$th scoring iteration for $\beta$, in the class of linear and nonlinear regression model, can be written as   
$
	\beta^{(m+1)}= \beta^{(m)}+ (J^\top_1 \Phi^{(m)} W^{(m)}J_1)^{-1} J^\top_1 \Phi^{(m)} T^{(m)} (y^*-\mu^{*(m)})
$,
where the $t$th elements of the vectors $y^*$ and $\mu^*$ are given in \eqref{eq3}. It is possible rewrite this equation in terms of weighted least squares
estimator as
$\beta^{(m+1)} =(J_1^{\!\top}\Phi^{(m)}W^{(m)}J_1)^{-1}\Phi^{(m)} J_1^{\!\top}W^{(m)}u_1^{(m)}.
$Here, 
$u_1^{(m)} =J_1 \beta^{(m)}+ {W^{-1}}^{(m)}T^{(m)}(y^*-{\mu^*}^{(m)})$.
Upon convergence, 
\begin{equation}\label{A5}
	\begin{split}
	\,\widehat{\!\beta} =(J_1^{\!\top}\,\widehat{\! \Phi}\,\widehat{\! W}J_1)^{-1} \,\widehat{\! \Phi}J_1^{\!\top}\,\widehat{\! W}u_1 \quad\text{where}\quad
		u_1=J_1{\,\widehat{\! \beta}} + {\,\widehat{\! W}}^{-1}{\,\widehat{\! T}}(y^*-{\widehat{{\mu}}^*}). \end{split} 
\end{equation}
Here, $\widehat{W}$, $\widehat{T}$, $\widehat{H}$ and $\widehat{D}$ are the matrices $W$, $T$, $H$ and $D$, respectively, evaluated at the maximum likelihood estimates. 
We note that $\,\widehat{\!\beta}$  in \eqref{A5} can be viewed as the least squares estimates of $\beta$ obtained by regressing $\,\widehat{\Phi}^{1/2} \,\widehat{\! W}^{1/2}u_1$ on $\,\widehat{\Phi}^{1/2}\,\widehat{\! W}^{1/2}J_1$.

\bibliographystyle{chicago}
%\bibliographystyle{rss}
%\bibliographystyle{splncs03}
%\bibliography{PRESS}

\begin{thebibliography}{}

\bibitem[\protect\citeauthoryear{Akaike}{Akaike}{1973}]{Aka:1973}
Akaike, H. (1973).
\newblock Information theory and an extension of the maximum likelihood
  principle.
\newblock In {\em Second International Symposium on Information Theory
  (Tsahkadsor, 1971)}, pp.\  267--281. Budapest: Akad\'emiai Kiad\'o.

\bibitem[\protect\citeauthoryear{Allen}{Allen}{1974}]{All:1974}
Allen, D.~M. (1974).
\newblock The relationship between variable selection and data augmentation and
  a method for prediction.
\newblock {\em Technometrics\/}~{\em 16}, 125--127.

\bibitem[\protect\citeauthoryear{Bartoli}{Bartoli}{2009}]{Bartoli2009}
Bartoli, A. (2009).
\newblock On computing the prediction sum of squares statistic in linear least
  squares problems with multiple parameter or measurement sets.
\newblock {\em International Journal of Computer Vision\/}~{\em 85\/}(2),
  133--142.

\bibitem[\protect\citeauthoryear{Bayer and Cribari-Neto}{Bayer and
  Cribari-Neto}{2017}]{BAYER2017}
Bayer, F.~M. and F.~Cribari-Neto (2017).
\newblock Model selection criteria in beta regression with varying dispersion.
\newblock {\em Communications in Statistics, Simulation and Computation\/}~{\em
  46}, 720--746.

\bibitem[\protect\citeauthoryear{Brascum, Johnson, and Thurmond}{Brascum
  et~al.}{2007}]{Brascom+Johnson+Thurmond_2007}
Brascum, A.~J., E.~O. Johnson, and M.~C. Thurmond (2007).
\newblock Bayesian beta regression: applications to household expenditures and
  genetic distances between foot-and-mouth disease viruses.
\newblock {\em Australian and New Zealand Journal of Statistics\/}~{\em
  49\/}(3), 287--301.

\bibitem[\protect\citeauthoryear{Cepeda-Cuervo and Gamerman}{Cepeda-Cuervo and
  Gamerman}{2005}]{Cepeda+Gamerman_2005}
Cepeda-Cuervo, E. and D.~Gamerman (2005).
\newblock Bayesian methodoly for modeling parameters in the two parameter
  exponential family.
\newblock {\em Estad\'{\i}stica\/}~{\em 57}, 93--105.

\bibitem[\protect\citeauthoryear{Chien}{Chien}{2011}]{Chien2011}
Chien, L.-C. (2011).
\newblock Diagnostic plots in beta-regression models.
\newblock {\em Journal of Applied Statistics\/}~{\em 38\/}(8), 1607--1622.

\bibitem[\protect\citeauthoryear{Cook and Weisberg}{Cook and
  Weisberg}{1982}]{CookWeisberg82book}
Cook, R.~D. and S.~Weisberg (1982).
\newblock {\em Residuals and Influence in Regression}.
\newblock Chapman and Hall.

\bibitem[\protect\citeauthoryear{Cribari-Neto and Zeileis}{Cribari-Neto and
  Zeileis}{2010}]{Cribari2010}
Cribari-Neto, F. and A.~Zeileis (2010).
\newblock Beta regression in {R}.
\newblock {\em Journal of Statistical Software\/}~{\em 34\/}(2), 1--24.

\bibitem[\protect\citeauthoryear{Espinheira, Ferrari, and
  Cribari-Neto}{Espinheira et~al.}{2008a}]{Espinheira2008}
Espinheira, P., S.~Ferrari, and F.~Cribari-Neto (2008a).
\newblock On beta regression residuals.
\newblock {\em Journal of Applied Statistics\/}~{\em 35\/}(4), 407--419.

\bibitem[\protect\citeauthoryear{Espinheira, Ferrari, and
  Cribari-Neto}{Espinheira et~al.}{2014}]{Espinheira2014}
Espinheira, P.~L., S.~Ferrari, and F.~Cribari-Neto (2014).
\newblock Bootstrap prediction intervals in beta regressions.
\newblock {\em Computational Statistics\/}~{\em 29\/}(5), 1263--1277.

\bibitem[\protect\citeauthoryear{Espinheira, Ferrari, and
  Cribari-Neto}{Espinheira et~al.}{2008b}]{Espinheira+Ferrari+Cribari_2008b}
Espinheira, P.~L., S.~L.~P. Ferrari, and F.~Cribari-Neto (2008b).
\newblock Influence diagnostics in beta regression.
\newblock {\em Computational Statistics and Data Analysis\/}~{\em 52},
  4417--4431.

\bibitem[\protect\citeauthoryear{Espinheira, Santos, and
  Cribari-Neto}{Espinheira et~al.}{2017}]{ESPINHEIRA2017}
Espinheira, P.~L., E.~G. Santos, and F.~Cribari-Neto (2017).
\newblock On nonlinear beta regression residuals.
\newblock {\em Biometrical Journal\/}~{\em n/a\/}(n/a), n/a--n/a.

\bibitem[\protect\citeauthoryear{Ferrari and Cribari-Neto}{Ferrari and
  Cribari-Neto}{2004}]{Ferrari2004}
Ferrari, S. and F.~Cribari-Neto (2004).
\newblock Beta regression for modelling rates and proportions.
\newblock {\em Journal of Applied Statistics\/}~{\em 31\/}(7), 799--815.

\bibitem[\protect\citeauthoryear{Figuero-Z{\'u}{\~n}iga, Arellano-Valle, and
  Ferrari}{Figuero-Z{\'u}{\~n}iga et~al.}{2013}]{Vale+Ferrari+Zuniga_2013}
Figuero-Z{\'u}{\~n}iga, J.~I., R.~B. Arellano-Valle, and S.~L. Ferrari (2013).
\newblock Mixed beta regression: A bayesian perspective.
\newblock {\em Computational Statistics and Data Analysis\/}~{\em 61},
  137--147.

\bibitem[\protect\citeauthoryear{McCullagh and Nelder}{McCullagh and
  Nelder}{1989}]{MCCULLAGH1989}
McCullagh, P. and J.~A. Nelder (1989).
\newblock {\em Generalized Linear Models\/} (2 ed.).
\newblock London: Chapman and Hall.

\bibitem[\protect\citeauthoryear{Mediavilla, F, and Shah}{Mediavilla
  et~al.}{2008}]{Mediavilla2008}
Mediavilla, F., L.~F, and V.~A. Shah (2008).
\newblock A comparison of the coefficient of predictive power, the coefficient
  of determination and aic for linear regression.
\newblock In K.~JE (Ed.), {\em Decision Sciences Institute, Atlanta}, pp.\
  1261--1266.

\bibitem[\protect\citeauthoryear{Nagelkerke}{Nagelkerke}{1991}]{Nagelkerke1991}
Nagelkerke, N. (1991).
\newblock A note on a general definition of the coefficient of determination.
\newblock {\em Biometrika\/}~{\em 78\/}(3), 691--692.

\bibitem[\protect\citeauthoryear{Ospina, Cribari-Neto, and Vasconcellos}{Ospina
  et~al.}{2006}]{Ospina2006960}
Ospina, R., F.~Cribari-Neto, and K.~L. Vasconcellos (2006).
\newblock Improved point and interval estimation for a beta regression model.
\newblock {\em Computational Statistics \& Data Analysis\/}~{\em 51\/}(2), 960
  -- 981.

\bibitem[\protect\citeauthoryear{Palmer and O'Connell}{Palmer and
  O'Connell}{2009}]{Palmer}
Palmer, P.~B. and D.~G. O'Connell (2009, sep).
\newblock Regression analysis for prediction: Understanding the process.
\newblock {\em Cardiopulmonary Physical Therapy Journal\/}~{\em 20\/}(3),
  23--26.

\bibitem[\protect\citeauthoryear{Pregibon}{Pregibon}{1981}]{pregibon1981}
Pregibon, D. (1981, 07).
\newblock Logistic regression diagnostics.
\newblock {\em The Annals of Statistics\/}~{\em 9\/}(4), 705--724.

\bibitem[\protect\citeauthoryear{Schwarz}{Schwarz}{1978}]{Schwarz78}
Schwarz, G. (1978).
\newblock {Estimating the dimension of a model}.
\newblock {\em Annals of Statistics\/}~{\em 6\/}(2), 461--464.

\bibitem[\protect\citeauthoryear{Simas, Barreto-Souza, and Rocha}{Simas
  et~al.}{2010}]{SIMAS2010}
Simas, A.~B., W.~Barreto-Souza, and A.~V. Rocha (2010).
\newblock Improved estimators for a general class of beta regression models.
\newblock {\em Computational Statistics \& Data Analysis\/}~{\em 54\/}(2),
  348--366.

\bibitem[\protect\citeauthoryear{Smithson and Verkuilen}{Smithson and
  Verkuilen}{2006}]{Smithson2006}
Smithson, M. and J.~Verkuilen (2006).
\newblock {A Better Lemon Squeezer? Maximum-Likelihood Regression With
  Beta-Distributed Dependent Variables}.
\newblock {\em Psychological Methods\/}~{\em 11\/}(1), 54--71.

\bibitem[\protect\citeauthoryear{Spiess and Neumeyer}{Spiess and
  Neumeyer}{2010}]{Spiess2010}
Spiess, A.-N. and N.~Neumeyer (2010).
\newblock An evaluation of $r^2$ as an inadequate measure for nonlinear models
  in pharmacological and biochemical research: a monte carlo approach.
\newblock {\em BMC Pharmacology\/}~{\em 10\/}(1), 6.

\bibitem[\protect\citeauthoryear{Zerbinatti}{Zerbinatti}{2008}]{Zerbinatti2008}
Zerbinatti, L. (2008).
\newblock Predi\c{c}\~{a}o de fator de simultaneidade atrav\'{e}s de modelos de
  regress\~{a}o para propor\c{c}\~{o}es cont\'{\i}nuas.
\newblock Msc thesis, University of S\~{a}o Paulo.

\end{thebibliography}

\end{document}